\documentclass[showpacs,superscriptaddress,twocolumn,prb,aps,footnoteinbib]{revtex4-1}
\usepackage[pdftex]{graphicx}
\usepackage{amsfonts}
\usepackage{amsmath}
\usepackage{amssymb}
\usepackage{upgreek}
\usepackage{dsfont}
\usepackage{bm}
\usepackage{color}
\usepackage[colorlinks=true,
            linkcolor=blue,
            urlcolor=blue,
            citecolor=blue]{hyperref}
\usepackage{times}
\newcommand{\calx}{\mathcal{X}}
\newcommand{\caly}{\mathcal{Y}}
\newcommand{\calz}{\mathcal{Z}}

\begin{document}
\title{Transport signatures of surface potentials on three-dimensional topological insulators}
\author{Sthitadhi Roy}
\affiliation{Max-Planck-Institut f\"ur Physik komplexer Systeme, N\"othnitzer 
Stra{\ss}e 38, 01187 Dresden, Germany}

\author{Sourin Das}
\affiliation{Department of Physics and Astrophysics,
University of Delhi, Delhi 110 007, India}

\begin{abstract}
The spin-momentum locked nature of the robust surface states of three dimensional topological insulators (3D TI) make them promising candidates for spintronics applications. Surface potentials which respect time reversal symmetry can exist at the surface between a 3D TI and the trivial vacuum. These potentials can distort the spin texture of the surface states while retaining their gapless nature. In this work, the effect of all such surface potentials on the spin textures is studied. Since, a tunnel magnetoresistance signal carries the information of the spin texture, it is proposed that spin-polarized tunneling of electrons to a 3D TI surface can be used to uniquely identify the surface potentials and quantitatively characterize them.
\end{abstract}

\maketitle

\section{Introduction \label{sec:intro}}
Topological insulators are exotic materials with an insulating bulk characterized by a gapped energy spectrum and topologically protected conducting surface states characterized by massless Dirac fermions \cite{Kane2005,Kane2005a,Bernevig2006,Bernevig2006a,Konig2007,Fu2007,Moore2010,Hasan2010,Qi2011,Bernevig2013book}.
The surface states of strong three dimensional topological insulators (3D TI)
\cite{Fu2007,Zhang2009,Chen2009,Moore2010,Hasan2010,Qi2011,Bernevig2013book} are robust against disorder (which respect time reversal) induced backscattering as they are topologically protected by time reversal symmetry \cite{Fu2007,Bernevig2013book,Moore2007,Roy2009,Zhang2009,Qi2010}.
These surface states are special with regard to them being perfectly spin-momentum locked. This feature of the surface states have been confirmed experimentally using spin resolved angle resolved photoemission spectroscopy (ARPES)\cite{Hsieh2008,Chen2009,Xia2009,Hasan2014}. Study of disorder induced quasi-particle interference patterns using a spin polarized scanning tunneling microscope (SP-STM) \cite{Roushan2009,Zhang2009a,Alpichshev2010} have also been used to probe this nature of surface states. Experiments with multiple ferromagnetic contacts have also been suggested \cite{Yokoyama2010,Taguchi2014} and they do show indications of spin-momentum locking \cite{Li2014a,Liu2014,Dankert2014}.
More recently, there have also been proposals\cite{Roy2015,Roy2015a} to reconstruct their spin texture of the surface states using spin polarized injection of currents from a source like an SP-STM and extracting an effective tunnel magnetoresistance\cite{Slonczewski1989} within a multi-terminal electrical transport setup.

The spin-momentum locked nature of the surface states provide a lot of scope for manipulating and directing spin polarized current in a controlled fashion on the surfaces of these materials. This along with robustness of these surface states make them the ideal materials for applications to spintronics\cite{Zutic2004,Yokoyama2014}. It then becomes extremely important to understand any other effect that can potentially distort the spin texture. Indeed, there can exist potentials localized on the surfaces of the 3D TIs which could affect the spin texture. It is then also desirable to have a protocol to characterize these surface potentials on a given sample.

In principle, both time reversal symmetry breaking potential and potentials which respects time reversal could be present on the surface of a topological insulators. Observation of robust surface state in experiments strongly indicate the absence of former but the presence of the later can not be ruled out in general. Hence studying the effect of such potentials can be of importance. In a recent work by  Zhang et al. (Ref.\onlinecite{Zhang2012}) an exhaustive list of such surface potentials was presented.  Zhang et al. studied the effect of these potentials to explain doping related effect observed in experiments on 3D TI surface states. In this article we will study the effect of such potentials on the spin texture of the surface states.

Electrons on a planar surface of a 3D TI can be described in terms of  two independent SU(2) degrees of freedom. For the surfaces perpendicular to the crystal growth are the electron spin and the orbital pseudospin. It turns out that these surface potentials can be classified into two categories, one that couples only to the orbital pseudospin of the electrons and the other which couples only to the spin. The second category of surface potentials which affect the spin textures is the focus of this work.

It turns out that these surface potentials have non-trivial effects on the surface states which lead to novel spin textures. For instance, some of the surface potentials tilt the spin out of the plane of the momentum leading to an imperfect spin-momentum locking. At some particular values of the surface potentials, the spin texture is distorted in such a way, that it looks as if it belongs to a Dresselhaus spin-orbit coupling even though the spin-orbit coupling term in the Hamiltonian governing the surface states is of the Rashba-type\cite{Liu2010}. Another interesting effect of the surface potentials found was, the spin texture gets twisted in the momentum-plane such that the spin-momentum locking stays perfect, albeit with a new spin-momentum locking angle. In this work, we study the effect of the surface potentials on the spin textures and provide analytical expressions for the spin textures in the presence of these potentials. 

The central message of this work is demonstrating how novel spin-textures can be obtained on the surface states of 3D TIs if the surface potentials can be manipulated in a controlled fashion. Further, using the protocol developed in Ref.\onlinecite{Roy2015} we describe how these new spin textures could be identified using simple quantum transport experiments. In fact, our proposal also provides a direct route to the quantitative identification of the surface potentials present on a sample.

The rest of the paper is organized as follows. In Sec.\ref{sec:potentials}, the effect of the surface potentials on the spin textures of the surface states is described.  Sec.\ref{sec:tunn} briefly sketches the connection between spin-polarized tunneling current and the spin texture. Sec.\ref{sec:sign} demonstrates how spin-polarized tunneling can be used to extract information about the surface potentials via the spin textures and finally Sec.\ref{sec:discuss} discusses experimental feasibilities and summarizes the results.

%%%%%%%%%%%%%%%%%%%%%%%%%%%%%%%%%%%%%%%%%%%%%%%%%%%
%%%%%%%%%%%%%%%%%%%%%%%%%%%%%%%%%%%%%%%%%%%%%%%%%%%
%%%%%%%%%%%%%%%%%%%%%%%%%%%%%%%%%%%%%%%%%%%%%%%%%%%
%%%%%%%%%%%%%%%%%%%%%%%%%%%%%%%%%%%%%%%%%%%%%%%%%%%
\section{Effect of surface potentials on spin texture of surface states \label{sec:potentials}}

Electrons in 3D TI materials like $\mathrm{Bi_2Se_3}$, $\mathrm{Bi_2Te_3}$ etc. have two SU(2) degrees of freedom, the orbital ($\bm{\tau}$) and the spin ($\bm{\sigma}$). The low energy physics of these materials is captured by the Hamiltonian
\begin{equation}
\mathcal{H}_{bulk}=  \mathds{I}_{2}\otimes (-m_0\tau_z + v_zk_z\tau_y)  + v_{\|}(k_y\sigma_x - k_x\sigma_y)\otimes \tau_x 
\label{eq:ham_bulk}
\end{equation}
where it is assumed that the crystal growth axis of the sample is along the $z$-direction. The boundaries of these crystals host topologically protected surface states as the mass undergoes as inversion in sign across the boundary from a 3D TI to the trivial vacuum. These surface states are evanescent on both the sides of the boundary as dictated by the Jackiw-Rebbi solution \cite{Jackiw1976} for the one-dimensional quantum mechanical problem along a direction perpendicular to the surface. To obtain these surface states, one effectively needs to calculate the evanescent states at the boundary between the TI and the vacuum at the $z=0$ plane, using the Hamiltonian in Eq.\ref{eq:ham_bulk} using a positive mass $m_0= m$ on the TI side (say $z<0$) and a negative mass $m_0=-M$ on the vacuum side ($z>0$). Since, the vacuum is a trivial insulator, the limit of $M\rightarrow\infty$ is taken. 

The surface states so obtained are described by a massless Dirac Hamiltonian of the form 
\begin{equation}
\mathcal{H}_{surf} = v_{\|}(\sigma^x k_y - \sigma^y k_x),
\label{eq:ham_surf}
\end{equation}
whose eigenstates are perfectly spin-momentum locked. The Jackiw-Rebbi solution also says that the surface state on the $z=0$ surface is an eigenstate of the $\tau^x$ operator with eigenvalue $+1$\cite{Zhang2012}. The continuity of the solution at $z=0$ ensures that the spinor at $z=0^-$, also has the same structure as that of the TI surface except that the decay length scale of these states on the two sides are different as dictated by the Dirac mass. Hereafter we denote this spinor as $\vert\psi_0\rangle$ and the spin texture evaluated from it turns out to be 
\begin{equation}
\langle\bm{\sigma}\rangle=(\sin\theta_{\mathbf{k}},-\cos\theta_{\mathbf{k}},0).
\label{eq:spin0}
\end{equation}
where $\theta_{\mathbf{k}}=\tan^{-1}(k_y/k_x)$.

However, the presence of time reversal symmetric surface potentials can significantly affect the nature of the these surface states and consequently the spin texture. Note that, since the surface potentials respect time reversal symmetry, they do not destroy the surface states.

It is important to stress that the surface potential cannot simply be incorporated into the surface Hamiltonian in Eq.\ref{eq:ham_surf}. In fact, to study the effect of the surface potentials on the surface states, it is required to re-derive the surface states by solving the Jackiw-Rebbi problem across the mass inversion boundary, but crucially now in the presence of the surface potentials. Hence, the effective problem now, is to obtain the evanescent modes at $z=0$ of a system described by the Hamiltonian 
\begin{equation}
\mathcal{H} = \mathcal{H}_{bulk} + \mathcal{H}_{pot},
\label{eq:ham_tot}
\end{equation}
where 
\begin{equation}
\mathcal{H}_{pot} = \bm{\tilde\Delta}\cdot\bm{\sigma}\otimes\tau^y \delta(z),
\label{eq:ham_pot}
\end{equation}
with $\bm{\tilde\Delta} = (2v_z/m)\bm{\Delta}$ and $\bm{\Delta} = (\Delta_x,\Delta_y,\Delta_z)$.

The spinor on the trivial vacuum side stays the same as $\vert\psi_0\rangle$ because the limit of $M\rightarrow+\infty$ is anyway taken. However, the Dirac Hamiltonian being linear in the momentum operator ( first order spatial derivative), the surface potentials modeled using the Dirac-delta functions (Eq.\ref{eq:ham_pot}) lead to discontinuities in the spinor across the surface between the TI and the trivial vacuum.
 As a result, the surface state spinors are not described anymore by $\vert\psi_0\rangle$, rather, they are described by a different spinor, denoted by $\vert\psi_\Delta\rangle$. It can be shown that the spinor $\vert\psi_\Delta\rangle$ can be obtained from $\vert\psi_0\rangle$ via a matrix $\mathcal{M}$ (see Appendix \ref{sec:disc} for the derivation) as
\begin{equation}
\vert \psi_\Delta\rangle = \mathcal{M}\vert \psi_0\rangle,
\label{eq:psidelta}
\end{equation}
 which appropriately implements the topological boundary condition \cite{Zhang2012} and hence gives the discontinuities. The form of $\mathcal{M}$ is given by 
\begin{equation}
\mathcal{M}=\frac{1}{m^2+\vert\bm{\Delta}\vert^2}\left[(m^2-\vert\bm{\Delta}\vert^2)\mathds{I}_4 - 2i m \bm{\Delta}\cdot\bm{\sigma}\otimes\mathds{I}_2\right].
\label{eq:M}
\end{equation}
A closer inspection of Eq.\ref{eq:M} reveals that it can be written as a direct product of a rotation matrix in the $\sigma$ sector, and an identity matrix in the $\tau$ sector as
\begin{equation}
 \mathcal{M}=e^{-i\theta_\Delta(\bm{\hat{\Delta}}\cdot\bm{\sigma})}\otimes\mathds{I}_2;~~
 \theta_\Delta=\cos^{-1}\left(\frac{m^2-\vert\bm\Delta\vert^2}{m^2+\vert\bm\Delta\vert^2}\right)
\label{eq:Mrot}
 \end{equation}
Hence the effect of the surface potentials can be geometrically interpreted as a rotation of the spin of the electron by an angle $2\theta_\Delta$ about the axis $\bm{\hat{\Delta}}$, while not affecting the orbital pseudospin.
The identification of $\mathcal{M}$ as a rotation matrix in the $\sigma$ sector allows for a broad classification of the surface potentials into two classes, one which rotates the spin textures about an axis which lies in the plane of the surface \textit{i.e.} $\Delta_{x(y)}\ne0$, $\Delta_z=0$ and the other which rotates the spin textures about an axis perpendicular to the surface \textit{i.e.} $\Delta_{x(y)}=0$, $\Delta_z\ne0$. The surface potentials of the first class can actually tilt the spin texture out of the plane of the surface. The maximum angle by which it can rotate the spin is $\pi$ ($\vert \bm{\Delta}\vert = m$) when the spin texture is back to being a planar one again. The important feature to note here is that, after a rotation by $\pi$  the spin-texture looks as if it comes from a Dresselhaus spin-orbit coupled Hamiltonian. This extrema is indeed very special as it changes the winding number of the spin as one moves along the Fermi surface from $+1$ to $-1$. The other class of surface potential is fundamentally different as it always rotates the spin texture in the plane of the surface and hence can never change the winding number of the spin texture along the Fermi surface.
It is also interesting to note that in both the limits of $\vert\bm{\Delta}\vert/m\gg1$ and $\vert\bm{\Delta}\vert/m\ll1$ the angle $\theta_\Delta\rightarrow 0$ which points to the fact that there is an intermediate window in the magnitude of the surface potentials where it has a significant effect on the spin textures.
It is important to note that since the surface potentials imposed are in the form of a direct product between the $\sigma$ and $\tau$ spaces, the resultant $\mathcal{M}$ matrix is also of the same structure. Consequently the direct product form of the spinors (in the $\tau$ and the $\sigma$ sector) for the surface states are retained.
The spin texture in the presence of the surface potentials can be obtained by taking an expectation value of the $\bm{\sigma}$ operator as $\langle\psi_\Delta\vert\bm{\sigma}\vert\psi_\Delta\rangle$. The distorted spin textures turn out to be of the form
\begin{subequations}
\begin{equation}
\langle\sigma^x\rangle = \calx_c\cos\theta_{\mathbf{k}} + \calx_s\sin\theta_\mathbf{k},
\end{equation}
\begin{equation}
\langle\sigma^y\rangle = \caly_c\cos\theta_{\mathbf{k}} + \caly_s\sin\theta_\mathbf{k},
\end{equation}
\begin{equation}
\langle\sigma^z\rangle = \calz_c\cos\theta_{\mathbf{k}} + \calz_s\sin\theta_\mathbf{k},
\end{equation}
\label{eq:sdelta}
\end{subequations}
where the coefficients of $\cos\theta_\mathbf{k}$ and $\sin\theta_\mathbf{k}$ are functions of $\mathbf{\Delta}$ given by
\begin{widetext}
\begin{subequations}
\begin{equation}
\calx_c = -\frac{4 ({\Delta^\prime_z} ({\Delta^\prime_x}^2+{\Delta^\prime_y}^2-1)+2 {\Delta^\prime_x} {\Delta^\prime_y}+{\Delta^\prime_z}^3)}{(1+\vert\mathbf{\Delta^\prime}\vert^2)^2};~~~~~~
\calx_s = 1-\frac{8(\Delta_y^{\prime 2}+\Delta_z^{\prime 2})}{(1+\vert\mathbf{\Delta^\prime}\vert^2)^2}
\end{equation}
\begin{equation}
\caly_c = -1 + \frac{8(\Delta_x^{\prime 2}+\Delta_z^{\prime 2})}{(1+\vert\mathbf{\Delta^\prime}\vert^2)^2};~~~~~~
\caly_s = -\frac{4 ({\Delta^\prime_z}( {\Delta^\prime_x}^2+{\Delta^\prime_y}^2-1)+2 {\Delta^\prime_x} {\Delta^\prime_y}+{\Delta^\prime_z}^3)}{(1+\vert\mathbf{\Delta^\prime}\vert^2)^2}\end{equation}
\begin{equation}
\calz_c = \frac{4(\Delta_x^{\prime 3} - 2\Delta_y^\prime\Delta_z^\prime+\Delta_x^\prime(-1+\Delta_y^{\prime 2}+\Delta_z^{\prime 2}))}{(1+\vert\mathbf{\Delta^\prime}\vert^2)^2};~~~~~~
\calz_s = \frac{4(\Delta_y^{\prime 3} + 2\Delta_x^\prime\Delta_z^\prime+\Delta_y^\prime(-1+\Delta_x^{\prime 2}+\Delta_z^{\prime 2}))}{(1+\vert\mathbf{\Delta^\prime}\vert^2)^2}
\end{equation}
\label{eq:spindelta}
\end{subequations}
\end{widetext}
where the $\Delta^\prime_i$s are dimensionless numbers given by $\Delta_i/m$.

The effect of the different surface potentials can be understood by looking at Eq.\ref{eq:spindelta} for the different cases separately. Firstly, in the absence of any surface potential ($\mathbf{\Delta}=0$), Eq.\ref{eq:sdelta} and Eq.\ref{eq:spindelta} reproduce the pristine spin texture mentioned in Eq.\ref{eq:spin0}. However, the presence of the surface potentials indeed leads to distorted spin textures. 

By setting $\Delta_x^\prime\ne0$ and $\Delta_y^\prime=0=\Delta_z^\prime$ in Eq.\ref{eq:spindelta} the effect of the surface potential of the form $\Delta_x\sigma^x$ can be understood. It turns out that such a potential leaves the spin texture along the $x$-direction ($\langle\sigma^x\rangle$) undisturbed as $\calx_c=0$ and $\calx_s=1$ in this case. On the contrary, such a potential reduces the magnitude of the spin texture along the $y$-direction and also tilts the spin out of the $x$-$y$ plane which is evident from $\langle\sigma^z\rangle$ being non-zero.
This is exactly the effect that a rotation of the spin texture about the $\hat x$ axis produces.
The effect of surface potentials of the form $\Delta_y\sigma^y$ and $\Delta_z\sigma^z$ can also be understood similarly from the interpretation of the rotation of the spin texture as mentioned in Eq.\ref{eq:Mrot}.
%

%%%%%%%%%%%%%%%%%%%%%%%%%%%%%%%%%%%%%%%%%%%%%%%%%%% 
%%%%%%%%%%%%%%%%%%%%%%%%%%%%%%%%%%%%%%%%%%%%%%%%%%%
%%%%%%%%%%%%%%%%%%%%%%%%%%%%%%%%%%%%%%%%%%%%%%%%%%%
%%%%%%%%%%%%%%%%%%%%%%%%%%%%%%%%%%%%%%%%%%%%%%%%%%%

\section{Spin polarized tunneling \label{sec:tunn}}

In this section, the calculation of spin-polarized tunneling current on the surface of a 3D TI is outlined and shown that how it carries information of the spin texture on the Fermi surface of the 3D TI surface states. 
In the second quantized language, the ground states of a 3D TI and a spin-polarized electron source can be written as $\prod_{k\le k_F}\psi_{TI}(\mathbf{k})c_{\mathbf{k}}^\dagger$ and $\prod_{k\le k_F}\psi_{S}d_{\mathbf{k}}^\dagger$ respectively where the $\psi_{TI(S)}$ are the spinors describing the internal degrees of freedom and $c^\dagger(d^\dagger)$  are usual fermionic creation operators for the 3D TI surface (spin polarized source).
For a particular momentum mode, the spin of the 3D TI surface state and the spin polarized source can respectively be calculated as $\mathbf{S}_{TI}(\mathbf{k}) = \langle\psi_{TI}(\mathbf{k})\vert\bm{\sigma}\vert\psi_{TI}(\mathbf{k})\rangle$ and $\mathbf{S}_{S} = \langle\psi_{S}\vert\bm{\sigma}\vert\psi_{S}\rangle$. Note that the spin of the spin-polarized source has no momentum dependence.

The tunneling Hamiltonian between such a source and the 3D TI surface can be written as
\begin{equation}
 \mathcal{H}_{tunn} = J~(\Psi_S^{\dagger}(\mathbf{r}=0)\Psi_{TI}(\mathbf{r}=0) + \text{h.c}).
  \label{eq:tunnel}
\end{equation}
In the momentum space $\mathcal{H}_{tunn}$ looks like
\begin{equation}
 \mathcal{H}_{{tunn}} = J\sum_{\mathbf{k},k^\prime}(z_{\mathbf{k}} c_{\mathbf{k}}^{\dagger}d_{k^\prime} + \text{h.c}),
\end{equation}
where $J$ is the tunneling amplitude, $z_{\mathbf{k}}$ is the overlap of the spinors of the 3D TI surface state and the spin polarized source.
The overlap of the spinors has a clear geometrical interpretation as
\begin{equation}
\vert z_\mathbf{k}\vert^2 = \frac{1}{2}(1+\mathbf{S}_{TI}(\mathbf{k})\cdot\mathbf{S}_S)
\end{equation}
where the $\mathbf{S}_{TI}(\mathbf{k})$ and $\mathbf{S}_S$ are unit vectors representing the spin expectation values of the 3D TI surface state and the spin polarized source. Note that this is the well known tunnel magnetoresistance response form for transport between two spin-polarized materials \cite{Slonczewski1989}.

\begin{figure}
\begin{center}
\includegraphics[width=0.99\columnwidth]{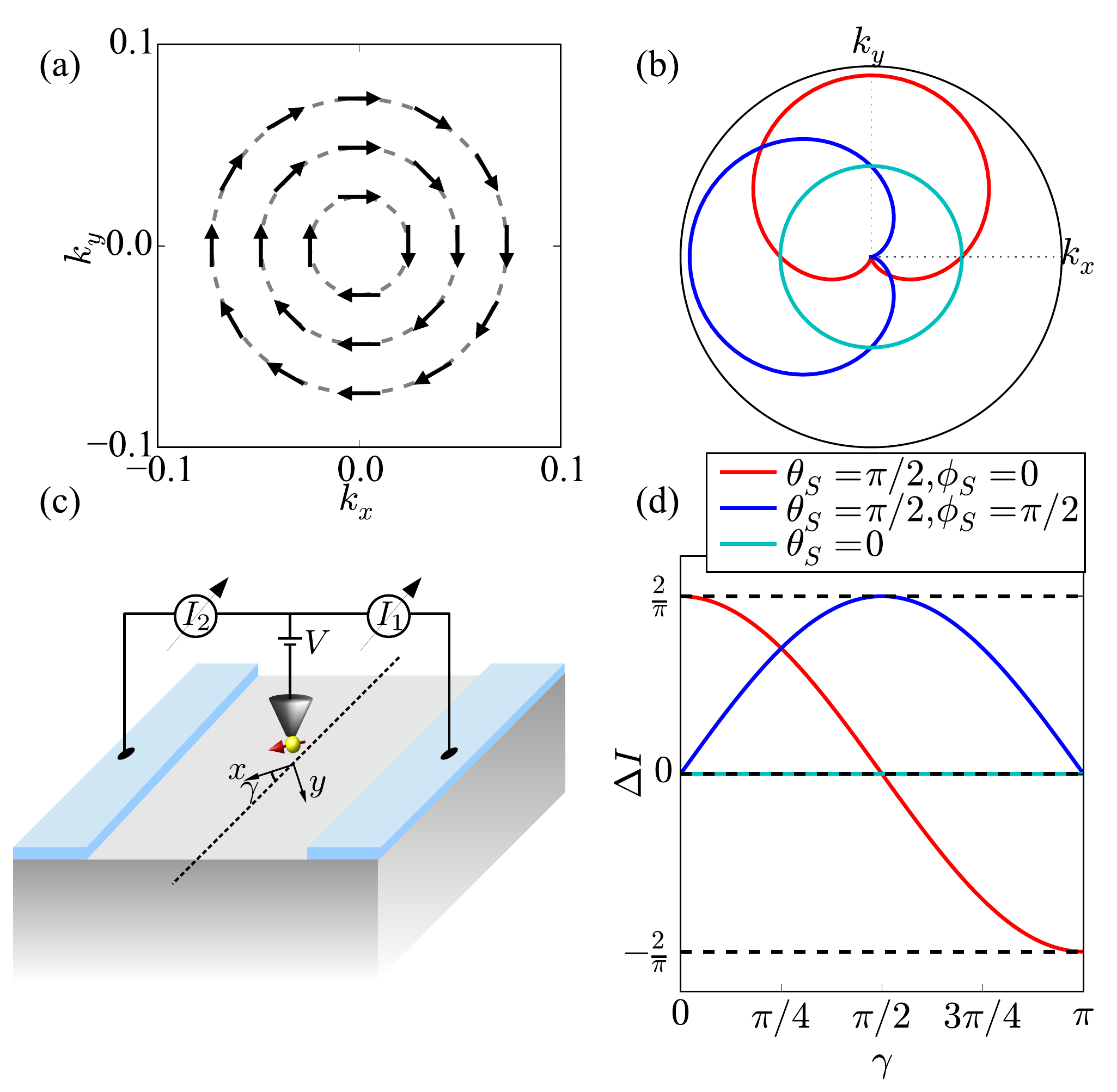}
\end{center}
\caption{(a) The spin texture of the pristine 3D TI surface is shown by the vectors. The circular dotted lines represent the Fermi surfaces and the momenta are represented in units of $\mathrm{\AA}^{-1}$. (b) The corresponding current distribution is shown in the momentum space for the injected electron having its spin along the $x$ (red), $y$ (blue) and the $z$ (cyan) direction. (c) A schematic of the proposed setup to measure the current asymmetry is shown. The symmetric line (dashed) about which the two contacts (shown in blue) are placed makes an angle $\gamma$ with the $x$ axis as shown. The current $I_1$ is the total current carried by momentum modes which have their azimuthal angles between $\gamma$ and $\gamma+\pi$, and current $I_2$ is carried by momentum modes whose azimuthal angles lie between $\gamma+\pi$ to $\gamma+2\pi$. The current asymmetry is defined as $\Delta I = I_1-I_2$. (d) The current asymmetries are plotted as a function of $\gamma$ for the three different spin directions of the injected electrons.}
\label{fig:curasym0}
\end{figure}

The total current turns out to be a sum of the momentum-resolved current values over the available momentum modes in the bias window \cite{Roy2015,Roy2015a}. This fact is extremely crucial as the decomposition of the total current into momentum-resolved currents enables us to divide the current that flows in each half of the TI surface to each of the contacts shown in Fig.\ref{fig:curasym0}(c) by simply summing the current over the momentum modes which go towards each contact from the spatial point $\mathbf{r}=0$. 
Presence of time reversal symmetry ensures that the net magnetization of the entire Fermi surface is identically zero, hence a current response summed over the entire Fermi surface cannot give any information about the spin texture. However, finite segments of the Fermi surface do have non-zero magnetization leading to a spin dependent response to tunneling which carry information about the spin texture, hence being able to separate the current response from segments of the Fermi surface is necessary to extract any information of the spin texture.

The expression for the momentum resolved current is given by \cite{Roy2015,Roy2015a}
\begin{eqnarray}                 
 I(\mathbf{k})=&& \frac{e J^2 \rho^{S}}{\hbar }(1+\mathbf{S}_{TI}(\mathbf{k})\cdot\mathbf{S}_S)\times \nonumber\\
                  &&(n_{\text{F}}(E_{\mathbf{k}},\mu_{TI},T_{TI}) -  n_{\text{F}}(E_{\mathbf{k}},\mu_{S},T_{S})),
                           \label{eq:Ik}
                           \end{eqnarray}
where $\rho^S$ is the density of the states of the spin-polarized source. The momentum resolve current allows us to define a dimensionless current symmetry (denoted by $\Delta I$) which filters out the information of the spin texture. For a given configuration of the contacts determined by the angle $\gamma$ (see Fig.\ref{fig:curasym0}c), $\Delta I$ is defined as
\begin{equation}
\Delta I= \frac{ \left(\int\limits_{\gamma}^{\gamma+\pi} - \int\limits_{\gamma+\pi}^{\gamma+2\pi}\right)d\theta_{\mathbf{k}}~\int\limits_0^\infty dk ~k ~I(\mathbf{k})}{ \int\limits_{0}^{2\pi}d\theta_{\mathbf{k}}~\int\limits_0^\infty dk ~k~ I(\mathbf{k})}.
\end{equation}

The way $\Delta I$ is defined makes it couple to the magnetization of half of the Fermi surface. Time reversal symmetry ensures that the magnetization of one half of the Fermi surface is always equal and opposite to that of the other half. The isotropy of the Fermi surface of the surface states of 3D TI tell us that the spin texture depends just on $\theta_{\mathbf{k}}$ and the Fermi energy just on $\vert\mathbf{k}\vert$. As a result, the current asymmetry $\Delta I$ takes a simple form given by
\begin{equation}
\Delta I  = \frac{2}{\pi}\mathbf{S}_S\cdot\int_\gamma^{\gamma+\pi} d\theta_{\mathbf{k}} ~\mathbf{S}_{TI}(\theta_\mathbf{k}) = \frac{2}{\pi}\mathbf{S}_S\cdot\mathbf{S}_{TI,half}(\gamma),
\label{eq:curasym}
\end{equation}
where $\mathbf{S}_{TI,half}(\gamma)$ denotes the magnetization of half of the Fermi surface between the azimuthal angles $\gamma$ and $\gamma+\pi$.

To illustrate how the dimensionless current asymmetry defined in Eq.\ref{eq:curasym} reconstructs the spin texture \cite{Roy2015},  the example of the spin texture of a pristine surface state (Eq.\ref{eq:spin0}) is shown in Fig.\ref{fig:curasym0}. The spin of the spin-polarized source can be parametrized using its polar angles $\mathbf{S}_S = (\sin\theta_S\cos\phi_S,\sin\theta_S\sin\phi_S,\cos\theta_S)$. As mentioned in Ref.\onlinecite{Roy2015}, the spin-momentum locking angle, $\theta_L$ for a planar spin texture can be obtained by setting $\theta_S=\pi/2$ and observing the angle $\gamma_{max}$ at which $\Delta I$ shows an extrema as $\gamma_{max} = \phi_S-\theta_L + \pi/2$. Note that the same magnitude of $\Delta I$ at $\gamma_{max}$ for $\phi_S=0$ and $\phi_S=\pi/2$ in Fig.\ref{fig:curasym0}c is a signature of the perfect spin-momentum locking. Also, the signs of the extrema tell us about the chirality and type of spin-orbit coupling, Rashba or Dresselhaus. In order to detect and reconstruct an out of plane spin polarization in the TI surface state, one simply needs to set $\theta_S=0$.

%%%%%%%%%%%%%%%%%%%%%%%%%%%%%%%%%%%%%%%%%%%%%%%%%%%
%%%%%%%%%%%%%%%%%%%%%%%%%%%%%%%%%%%%%%%%%%%%%%%%%%%
%%%%%%%%%%%%%%%%%%%%%%%%%%%%%%%%%%%%%%%%%%%%%%%%%%%
%%%%%%%%%%%%%%%%%%%%%%%%%%%%%%%%%%%%%%%%%%%%%%%%%%%

\section{Signatures of surface potentials in spin-polarized tunneling \label{sec:sign}}
The current asymmetry measurements carry signatures of the spin textures as discussed in Sec.\ref{sec:tunn} and the spin textures can carry signatures of the surface potentials as mentioned in Sec.\ref{sec:potentials}. Hence, a natural question is can the current asymmetry measurements directly characterize the surface potentials and this section discusses this question in details. It turns out that the form of the spin textures mentioned in Eq.\ref{eq:sdelta} leads to an easy identification of the functions in Eqs.\ref{eq:spindelta} via the current asymmetries and it is indeed possible to fully characterize the surface potentials from them.

The three components of the spin texture can be uniquely identified by three scans of the current asymmetries over $\gamma$, each with the spin of the electron from spin polarized source along three orthogonal directions. From the form of the spin textures in Eq.\ref{eq:sdelta}, it can be seen that the current asymmetries as a function of $\gamma$ for the injected electron spin being along $\hat x$, $\hat y$, and $\hat z$ are given by 
\begin{subequations}
\begin{equation}
\Delta I^{(x)} = \frac{2}{\pi}S^x_{TI,half}(\gamma)=\frac{2}{\pi}(-\mathcal{X}_c \sin\gamma + \mathcal{X}_s \cos\gamma)
\end{equation}
\begin{equation}
\Delta I^{(y)} =\frac{2}{\pi}S^y_{TI,half}(\gamma)= \frac{2}{\pi}(-\mathcal{Y}_c \sin\gamma + \mathcal{Y}_s \cos\gamma)
\end{equation}
\begin{equation}
\Delta I^{(z)} =\frac{2}{\pi}S^z_{TI,half}(\gamma)= \frac{2}{\pi}(-\mathcal{Z}_c \sin\gamma + \mathcal{Z}_s \cos\gamma)
\end{equation}
\end{subequations}
Hence, in an experiment one needs to fit the functions in Eq.\ref{eq:spindelta} to the experimentally measured current asymmetries to reconstruct the surface potentials present.

In order to understand the effect of the surface potentials on the current asymmetries, the classes of surface potentials, one which rotate the spin texture out of the plane, and the other which rotates the spin texture within the plane are discussed separately. Finally how the current asymmetries for a generic case with all surface potentials present with different magnitudes can reconstruct the surface potential configuration is discussed.

%
%%%%%%%%%%%%%%%%%%%%%%%%%%%%%%%%%%%%%%%%%%%%%%%%%%%%
\subsection{Surface potentials which rotate the spin out of the plane \label{sec:deltax}}
The spin textures are rotated out of the plane of the surface if the axis of rotation, $\bm{\hat\Delta}$ lies in the plane (in this case, the $x$-$y$ plane); the surface potentials of the form $\Delta_x\neq0$, $\Delta_y\neq0$, and $\Delta_z=0$ fall under this category. The component of the in-plane spin texture along $\bm{\hat\Delta}$ stays the same, however the in-plane component perpendicular to it gets reduced in magnitude. Accordingly, the out of plane component of the texture picks up a finite value.

\begin{figure}
\begin{center}
\includegraphics[width=0.99\columnwidth]{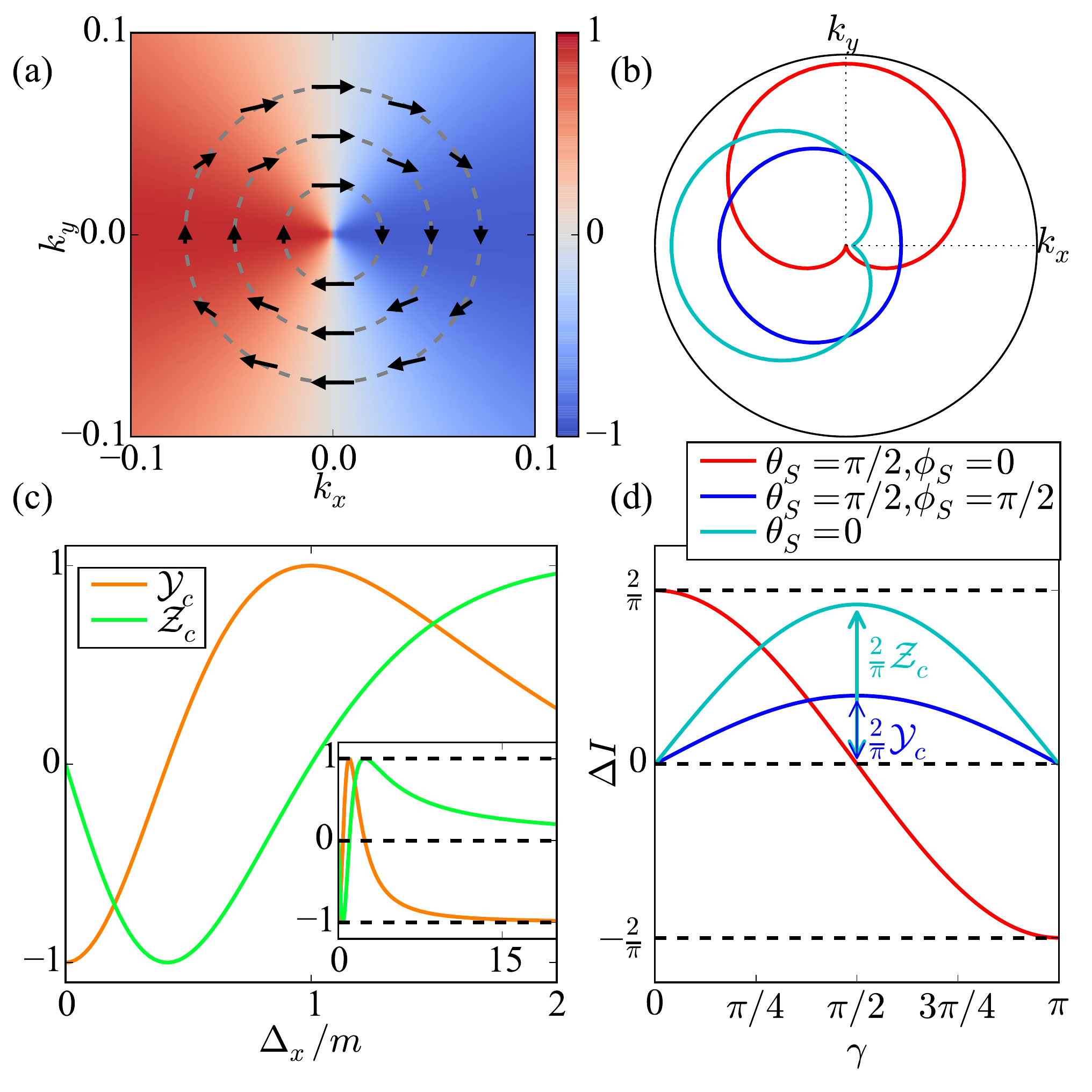}
\end{center}
\caption{(a) The spin texture is shown for $\Delta_x=0.3m, \Delta_y=0=\Delta_z$ with the vectors showing the in-plane spin textures $\langle\sigma^x\rangle$ and $\langle\sigma^y\rangle$ and the color density plot in the background showing the out of plane spin texture $\langle\sigma^z\rangle$. Note that the magnetization along the $y$ direction is reduced in magnitude. (b) For the same $\bm{\Delta}$, the current distribution is shown as a polar plot for three different spin directions of the injected electron. It can be seen that the maxima of the current distribution for the case of the injected spin along $y$ direction is reduced. Also, due to the spin texture along the $z$ direction, the current distribution for the injected spin along the $z$ direction is no longer circular. These facts are also reflected in the current asymmetries shown in (d). The behavior of the functions $\caly_c$ and $\calz_c$ as a function of $\Delta_x$ is shown in (c). The inset shows their behavior in the limit of large $\Delta_x$ where they tend to their pristine values.}
\label{fig:deltax}
\end{figure}

Owing to the azimuthal symmetry of the Fermi surface, without loss of generality we can analyze the case of $\Delta_x\ne0,\Delta_y=0=\Delta_z$ which rotates the spin texture about the $x$-axis and hence does not affect the spin texture along the $x$-direction
This is mathematically manifested as $\mathcal{X}_c=0$ and $\mathcal{X}_s=1$ for this case. However the functions $\caly_c$ and $\caly_s$ deviate from their pristine values of -1 and 0 respectively, and so do $\calz_c$ and $\calz_s$ from 0. This means that the spin textures along the $y$ and $z$ direction are affected and this is also evident from the example shown in Fig.\ref{fig:deltax}a.
It is clear that the component of the spin along the $y$ direction for every momenta gets reduced and the out of plane $z$ component picks up a finite value. 

Since the surface potential does not affect $\langle\sigma^x\rangle$, $\Delta I$ measured with the injected spin along the $x$ direction ($\theta_S=\pi/2$,$\phi_S=0$) has the same behavior with $\gamma$ as the pristine case. However, as the component of the spin texture along the $y$ direction is reduced in magnitude, the value of $\Delta_I$ for the injected spin along $y$ direction ($\theta_S=\pi/2$,$\phi_S=\pi/2$) at its extrema is also correspondingly reduced as shown in Fig.\ref{fig:deltax}d. 
In fact, the extremum value for $\Delta I^{(y)}$ carries a direct information of the magnitude of the surface potential $\Delta_x$ as $\Delta I^{(y)}\vert_{\gamma=\pi/2} = 2\caly_c/\pi$. 
The out of plane magnetization $\langle\sigma^z\rangle$ also picks up a texture corresponding to the effect on the $\langle\sigma^y\rangle$ due to the surface potential.
Hence the current asymmetry with injected spin along the $z$ direction also shows an extrema at $\gamma=\pi/2$ whose value is given by $2\calz_c/\pi$. To avoid the systematic errors in an experiment it is better to look at the ratios $\Delta I^{(y)}\vert_{\gamma=\pi/2}/\Delta_I^{(x)}\vert_{\gamma=0}$ and $\Delta I^{(z)}\vert_{\gamma=\pi/2}/\Delta_I^{(x)}\vert_{\gamma=0}$, whose experimentally obtained values can be fitted to the curves of $\caly_c$ and $\calz_c$ shown in Fig.\ref{fig:deltax}c to uniquely determine the magnitude of the surface potential $\Delta_x$. 

At this point, it is important to mention some particular features of the behaviors of $\caly_c$ and $\calz_c$. Note that in the limit of $\Delta_x\gg m$, $\caly_c$ and $\calz_c$ tend to their values of the pristine texture i.e -1 and 0 respectively. This implies that in the limit $\Delta_x\gg m$, its effect on the spin textures is negligible. $\Delta_x=m$ is another special value where $2\theta_\Delta = \pi$ which implies $\calz_c=0$ (its pristine value), whereas $\caly_c=1$ which is negative of its pristine value. This leads to the spin texture looking like as if it is obtained from a Dresselhaus spin-orbit coupling. A change of sign in $\Delta I^{(y)}$ as a function of $\gamma$ compared to the pristine case signals this. %
For $\Delta_x = (\sqrt{2}\pm1)m$, it turns out that $2\theta_\Delta=\pi/2$ which means the spin texture lies completely in the $x$-$z$ plane as manifested in $\caly_c$ = 0 and $\calz_c = \mp 1$. This situations would reflect itself in the current asymmetries as $\Delta I^{(y)}$ being identically zero for all $\gamma$ and $\vert(\Delta I^{(z)}\vert_{\gamma=\pi/2}/\Delta_I^{(x)}\vert_{\gamma=0})\vert$ taking its maximum value 1.

%
%%%%%%%%%%%%%%%%%%%%%%%%%%%%%%%%%%%%%%%%%%%%%%%%%%%%
%
\begin{figure}
\begin{center}
\includegraphics[width=0.99\columnwidth]{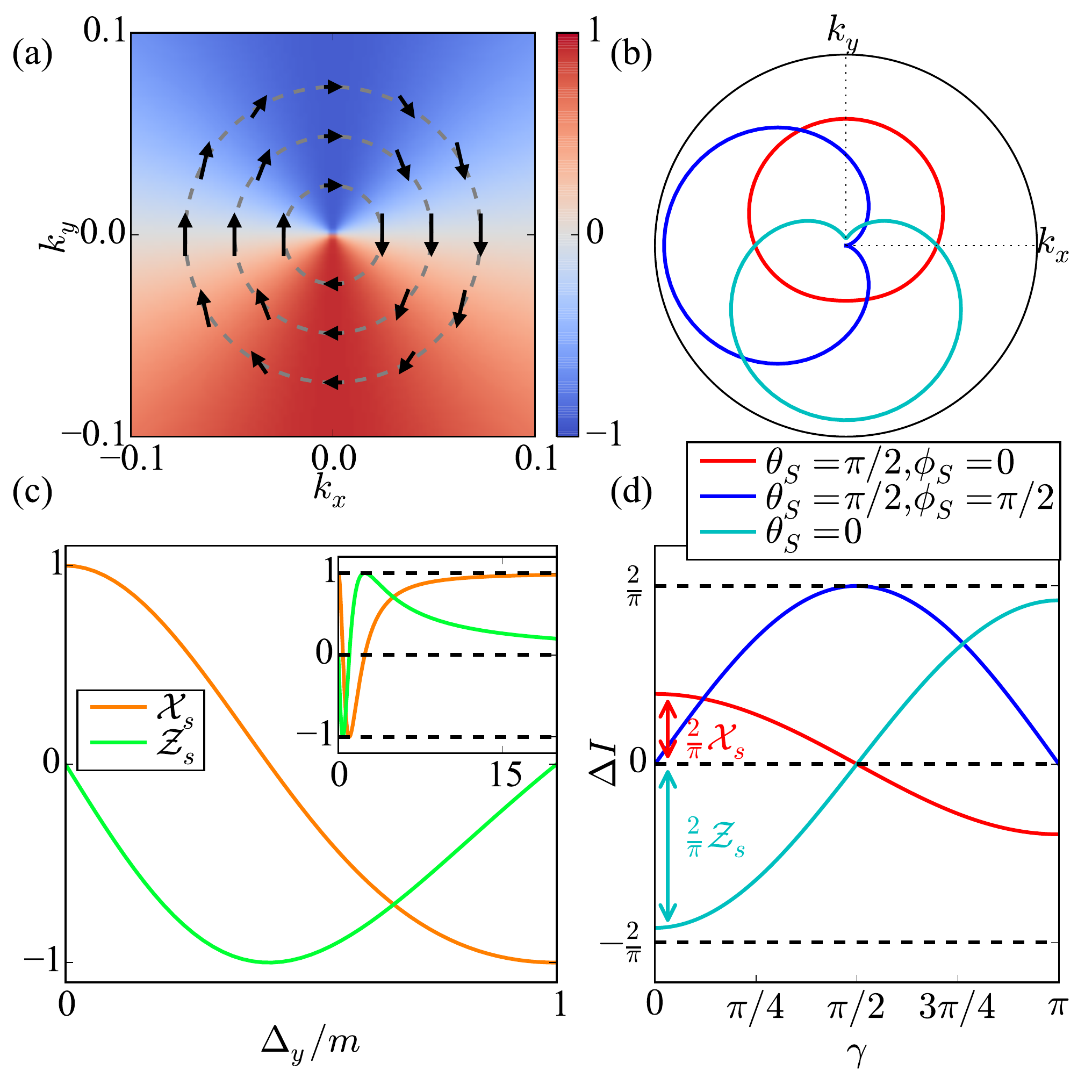}
\end{center}
\caption{(a) The spin texture is shown for $\Delta_y=0.3m, \Delta_x=0=\Delta_z$ with the vectors showing the in-plane spin textures $\langle\sigma^x\rangle$ and $\langle\sigma^y\rangle$ and the color density plot in the background showing the out of plane spin texture $\langle\sigma^z\rangle$. Note that the magnetization along the $x$ direction is reduced in magnitude. (b) For the same $\bm{\Delta}$, the current distribution is shown as a polar plot for three different spin directions of the injected electron. It can be seen that the maxima of the current distribution for the case of the injected spin along $x$ direction is correspondingly reduced. Also, due to the spin texture along the $z$ direction, the current distribution for the injected spin along the $z$ direction is no longer circular. These facts are also reflected in the current asymmetries shown in (d). The behavior of the functions $\calx_s$ and $\calz_s$ as a function of $\Delta_y$ is shown in (c). The inset shows their behavior in the limit of large $\Delta_y$ where they tend to their pristine values.}
\label{fig:deltay}
\end{figure}
For completeness we also present the results for the case of $\Delta_y\ne0,\Delta_x=0=\Delta_z$, which rotates the spin texture about the $y$-axis. As a result the spin texture along the $y$ direction remains unaffected and along the $x$ direction it reduces in magnitude. As before the out plane magnetization picks up a texture in accordance to the loss of magnetization along the $x$ direction.
It can be seen from Fig.\ref{fig:deltay}d that the current asymmetry profile for the injected spin along the $y$ direction is the same as the one for the pristine case which indicates that $\langle\sigma^y\rangle$ has the same texture as the pristine case i.e $\caly_c=-1$ and $\caly_s=0$.
However $\calx_s$ and $\calz_s$ deviate from their pristine values of $1$ and $0$ respectively, leading to $\Delta I^{(x)}$ and $\Delta I^{(z)}$ showing behaviors different from the pristine case. 
%
%
%
%%%%%%%%%%%%%%%%%%%%%%%%%%%%%%%%%%%%%%%%%%%%%%%%%%%%
\subsection{Surface potentials which rotate the spin in the plane\label{sec:deltaz}}
\begin{figure}
\begin{center}
\includegraphics[width=0.99\columnwidth]{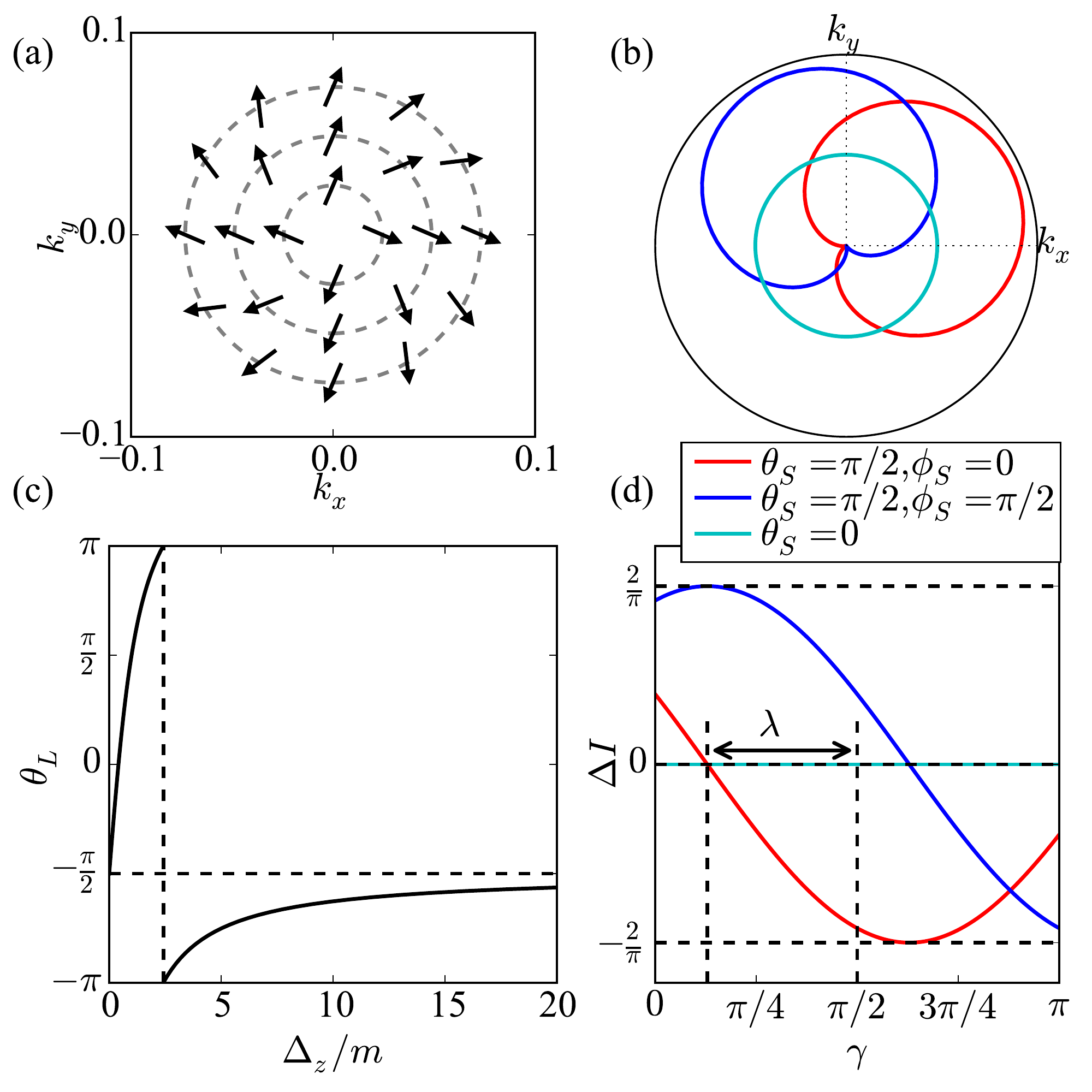}
\end{center}
\caption{(a) The twisted but in-plane spin textures for $\Delta_z=0.3m, \Delta_x=0=\Delta_y$ is shown. (b) The corresponding current distribution is shown as a polar plot. Note that the distribution compared to the pristine case shown in Fig.\ref{fig:curasym0} is rotated corresponding to the twist in the spin textures. (d) The twist in the spin texture is reflected in the current asymmetries as the profile is shifted in $\gamma$ by an angle $\lambda$ which gives the new spin-momentum locking angle.  (c) The spin-momentum locking angle is plotted as a function of the surface potential strength $\Delta_z$. Note that for large values of $\Delta_z$ the spin-momentum locking angle tends to its pristine value of $-\pi/2$.}
\label{fig:deltaz}
\end{figure}
The surface potential of the type $\Delta_z \sigma^z\otimes\tau^y$ has an effect on the spin texture which is quite different from the previous case as it rotates the spin for all the momenta by the angle $2\theta_\Delta$ about the $z$ axis and hence changes the resultant spin-momentum locking angle. Contrary to the previous case, it leaves the spin texture completely planar as it does not induce any out of plane magnetization on the surface states of the 3D TI.
The changed spin-momentum locking angle can be directly obtained from the current asymmetries, as for a particular spin direction of the injected electron, the extrema in $\Delta I$ occurs at $\gamma_{max} = \phi_S-\theta_L + \pi/2$. In the presence of the surface potentials the maximas of $\Delta I$ get shifted by angle $\lambda$ as shown in Fig.\ref{fig:deltaz}d. In an experiment the value of $\lambda$ can easily be obtained as mentioned and it could be used to directly obtain the value of the surface potential $\Delta_z$ by fitting $\lambda$ to $\sin^{-1}(\caly_s/\sqrt{\caly_c^2+\caly_s^2})$.

It is also interesting to note that at $\Delta_z = (\sqrt{2}\pm1)m$, $2\theta_\Delta=\pi/2$ which means the spin texture becomes completely radial with $\langle\bm\sigma\rangle = \pm\hat{r}$. This will be reflected by the $\Delta I$ as function of $\gamma$ getting shifted by $\pi/2$. The spin-momentum locking angle $\theta_L$ as a function of $\Delta_z$ is shown in Fig.\ref{fig:deltaz}c
%
%
%
%
%%%%%%%%%%%%%%%%%%%%%%%%%%%%%%%%%%%%%%%%%%%%%%%%%%%%
\subsection{Generic case}
In the previous two subsections the effect of the surface potentials which rotate the spin-texture in and out of the plane were studied separately. However, in a realistic generic case, it is expected that all of them would be present together with different magnitudes. In such a case, their effect on the spin texture could be a complicated combination of the effects discussed in Secs.\ref{sec:deltax} and \ref{sec:deltaz}. It is then imperative to uniquely extract the values $\Delta_x$, $\Delta_y$, and $\Delta_z$ from the current asymmetry measurements that are accessible experimentally.

It turns out that this is indeed possible via a constrained optimization with the constraints being
\begin{subequations}
\begin{equation}
\Delta I^{(x)}\vert_{\gamma=\pi/2} = -2\calx_c/\pi;~~~~ \Delta I^{(x)}\vert_{\gamma=0} = 2\calx_s/\pi,
\end{equation}
\begin{equation}
\Delta I^{(y)}\vert_{\gamma=\pi/2} = -2\caly_c/\pi;~~~~ \Delta I^{(y)}\vert_{\gamma=0} = 2\caly_s/\pi,
\end{equation}
\begin{equation}
\Delta I^{(z)}\vert_{\gamma=\pi/2} = -2\calz_c/\pi;~~~~ \Delta I^{(z)}\vert_{\gamma=0} = 2\calz_s/\pi.
\end{equation}
\label{eq:constraints}
\end{subequations}
Essentially this optimization is effectively a search for a triplet of values for $\Delta_x$, $\Delta_y$, and $\Delta_z$ which simultaneously satisfy the six constraints in Eq.\ref{eq:constraints}, which can be done by several numerical techniques.
\footnote{The optimization used by the authors to verify the protocol was implemented using the Broyden-Fletcher-Goldfarb-Shanno algorithm which gave relative errors of magnitudes less than $10^{-8}$. More details of the algorithm can be found in Ref.\onlinecite{Nocedal2006}}

%%%%%%%%%%%%%%%%%%%%%%%%%%%%%%%%%%%%%%%%%%%%%%%%%%%
%%%%%%%%%%%%%%%%%%%%%%%%%%%%%%%%%%%%%%%%%%%%%%%%%%%
%%%%%%%%%%%%%%%%%%%%%%%%%%%%%%%%%%%%%%%%%%%%%%%%%%%
%%%%%%%%%%%%%%%%%%%%%%%%%%%%%%%%%%%%%%%%%%%%%%%%%%%

\section{Discussions and Summary \label{sec:discuss}}
One of the crucial requirements for the practical implementation of our proposal is the ability to manipulate the spin-polarization of the tunneling tip. The outstanding progress in the field of direct observation of spin textures and magnetic order via spin-sensitive scanning probes like spin-polarized STM and magnetic exchange force microscopy indeed makes our proposal feasible (for a detailed review, see Ref.\onlinecite{Wiesendanger2009}). It was found that the polarization of spin-polarized tips prepared by coating non-magnetic tips (Tungsten (W)) with anti-ferromagnetic materials (Chromium (Cr)) can be manipulated by varying the thickness of the Cr layer.\cite{Wachowiak2002,Wiesendanger2009} For instance, W tips coated with 25-45 monolayers of Cr are sensitive to the out-of-plane polarization where as with ca.100 monolayers of Cr are sensitive to the in-plane polarization. \cite{Wachowiak2002,Wiesendanger2009,Yang2013} This provides a way of manipulating the polarization of the tip without changing its orientation to its crystalline axis or the crystal growth axis of the 3D TI. In our case, with the tip being sensitive to the out-of-plane polarization, the $z$ component of the spin texture can be measured. In order to scan the in-plane spin texture, the in-plane polarization of the tunneling tip can be fixed and one does not need to necessarily rotate it. The point to note here is that the tunnel magnetoresistance signal simply depends on the relative angle between the contact orientation and the tip polarization (Eq.\ref{eq:curasym} in the manuscript). By having multiple pairs of contacts with different orientations (denoted by the angle $\gamma$) on the same sample, the current asymmetry can be measured as a function of $\gamma$.

Note that in our case, with the in-plane polarization of tip fixed, one could also rotate the orientation of the contacts relative to the tip polarization, or alternatively, taking advantage of the azimuthal Fermi surface, rotate the sample with the contacts fixed so as to effectively change its orientation relative to the tip polarization.

The experimental feasibility of our proposals also relies on the ballistic nature of transport on the surface of the 3D TI. Hence, the elastic mean free path is a crucial length scale in the problem. Since the transport studied in the work is spin polarized, the spin relaxation length is a more relevant length scale, however in the absence of magnetic impurities it is bounded from below by the mean free path\cite{Pesin2012}.	The elastic mean free path of 3D TI material $\mathrm{Bi_2Te_3}$ has been reported to be 235nm \cite{Qu2010}, so it is expected that a sample with a micrometer-by-micrometer surface would be sufficient for the experiment and it is indeed a feasible sample size\cite{Roushan2009}. Our proposal also  relies on the contacts being reflectionless and as different amounts of reflections from the two contacts could lead to a skewed current asymmetry. However, the presence of time reversal symmetry and spin-momentum locking on the surfaces of the 3D TI ensure that the contacts are largely reflectionless as any reflection would require a spin flip which is not possible by non-magnetic contacts. In fact experiments with contact resistances as low as $\mathrm{1 m\Omega}$ have been reported \cite{Li2014a}.

In this work, the hexagonally warped regimes of the spectrum \cite{Fu2009} have not been explored, however the derivation of the matrix $\mathcal{M}$ is exactly the same in the presence of the warping term. So, the effect of the warping can be very simply taken into account by taking the appropriate form of $\vert\psi_0\rangle$ (in Eq.\ref{eq:psidelta}) to include its effect. In fact, the combination of the effects warping and surface potentials can lead to exotic spin textures on the surfaces.

It is also important to mention that though our calculations have been done for surfaces of 3D TIs perpendicular to the crystal growth axis, they can be straightforwardly be extended to surfaces at arbitrary angles ($\theta$) to the crystal growth via the transformations \cite{Zhang2012,Roy2015a} which take $\sigma\rightarrow S_\theta$ and $\tau \rightarrow T_\theta$ where $\bm{S}_\theta$ and $\bm{T}_\theta$ are $\theta$-dependent linear combinations of the spin and orbital pseudospin given by
\begin{equation}
\begin{split}
 \bm{T}_\theta&=\{\alpha\tau_x+\beta\tau_y\otimes \sigma_y,\alpha\tau_y -\beta\tau_x\otimes \sigma_y,\tau_z \}\\
 \bm{S}_\theta &= \{\alpha  \sigma_x-\beta\tau_z\otimes \sigma_z,\sigma_y, \alpha \sigma_z+\beta\tau_z\otimes\sigma_x\},
 \end{split}
\end{equation}
where $v_3=\sqrt{(v_{\|} \sin\theta)^2+(v_{z}\cos\theta)^2}$, $\alpha=v_z \cos\theta/v_3$, and  $\beta=v_{\|} \sin\theta/v_3$. 

It is also worth mentioning that our protocol can be extended to study weak disorder which varies slowly on the surface of the 3D TI provided the disorder potentials preserve time reversal symmetry, however a rigorous treatment of disorder on the surface of 3D TIs presents an interesting problem in itself.

Finally, to summarize the main results of the work, a systematic treatment of the all possible time reversal symmetric surface potentials on 3D TIs was presented and their effect on the spin-momentum locking of the surface states was studied. It was found that these surface potentials can have non-trivial effects leading to novel spin textures on the surfaces of 3D TIs. Quantum transport experimental protocols were suggested to study these spin textures using spin polarized electron injection into the surfaces of the TI leading to multi-terminal tunnel magnetoresistance  response. It was shown that these studies can be used to extract all the information quantitatively about the surface potentials, if present, on the surfaces of 3D TI and hence can act as a smoking gun for them. These findings are extremely important from the point of view of the spin-momentum locked surface states of 3D TIs being used for spintronics applications. Finally, the surface potentials can also serve as a platform for engineering novel spin textures on the surfaces of 3D TIs, provided a controlled way of manipulating the surface potentials in experiments is engineered.
%

%%%%%%%%%%%%%%%%%%%%%%%%%%%%%%%%%%%%%%%%%%%%%%%%%%%
%%%%%%%%%%%%%%%%%%%%%%%%%%%%%%%%%%%%%%%%%%%%%%%%%%%
%%%%%%%%%%%%%%%%%%%%%%%%%%%%%%%%%%%%%%%%%%%%%%%%%%%
%%%%%%%%%%%%%%%%%%%%%%%%%%%%%%%%%%%%%%%%%%%%%%%%%%%

\begin{acknowledgements}
The authors thank Eugene Mele for illuminating discussions regarding the incorporation of surface potentials on the surface states of 3D TIs and also thank Krishanu Roychowdhury for helpful discussions and critical comments on the manuscript.
\end{acknowledgements}

%%%%%%%%%%%%%%%%%%%%%%%%%%%%%%%%%%%%%%%%%%%%%%%%%%%
%%%%%%%%%%%%%%%%%%%%%%%%%%%%%%%%%%%%%%%%%%%%%%%%%%%
%%%%%%%%%%%%%%%%%%%%%%%%%%%%%%%%%%%%%%%%%%%%%%%%%%%
%%%%%%%%%%%%%%%%%%%%%%%%%%%%%%%%%%%%%%%%%%%%%%%%%%%
\onecolumngrid
\appendix*
\section{Derivation of the matrix $\mathcal{M}$ \label{sec:disc}}
In this appendix the derivation of the matrix $\mathcal{M}$ is sketched. The derivation involves taking the Hamiltonian with the surface potentials included and integrating the Dirac equation across the boundary between the 3D TI and the trivial vacuum to obtain the discontinuity in the spinors due to the surface potentials as they are modeled by a Dirac-delta function. As mentioned in the main text, the spinor on the trivial vacuum side ($z>0$) stays the same as $\vert\psi_0\rangle$ which we denote as $\vert\psi_0\rangle\vert_{z=0^+} = (a_0,b_0,c_0,d_0)^T$. The spinor on the TI side ($z<0$) is denoted by $\vert\psi_\Delta\rangle\vert_{z=0^-} =  (a_\Delta,b_\Delta,c_\Delta,d_\Delta)^T$

The Hamiltonian $\mathcal{H}$ in Eq.\ref{eq:ham_tot} in its matrix form is given by
\begin{equation}
\begin{pmatrix}
 -m_0 & - v_z \partial_z-i \tilde{\Delta}_z\delta(z) & 0 & \begin{smallmatrix}ike^{-i\theta_{\mathbf{k}}} v_{\|}-\\i \left(\tilde{\Delta}_x-i \tilde{\Delta}_y\right)\delta(z)\end{smallmatrix} \\
  v_z \partial_z+i \tilde{\Delta}_z\delta(z) & m_0 & \begin{smallmatrix}ike^{-i\theta_{\mathbf{k}}} v_{\|}+\\i \left(\tilde{\Delta}_x-i \tilde{\Delta}_y\right)\delta(z)\end{smallmatrix} & 0 \\
 0 &  \begin{smallmatrix}-ike^{i\theta_{\mathbf{k}}} v_{\|}-\\i \left(\tilde{\Delta}_x+i \tilde{\Delta}_y\right)\delta(z)\end{smallmatrix} & -m_0 & - v_z \partial_z + i \tilde{\Delta}_z\delta(z) \\
 \begin{smallmatrix}-ike^{i\theta_{\mathbf{k}}} v_{\|}+\\i \left(\tilde{\Delta}_x+i \tilde{\Delta}_y\right)\delta(z)\end{smallmatrix} & 0 &  v_z \partial_z-i \tilde{\Delta}_z\delta(z) & m_0 \\
\end{pmatrix},
\end{equation}

such that when the Dirac equation is integrated across the boundary at $z=0$ as
\begin{equation}
\lim_{\epsilon\rightarrow0}\int_{-\epsilon}^{+\epsilon}dz ~~ \mathcal{H}\vert\psi\rangle = \lim_{\epsilon\rightarrow0}\int_{-\epsilon}^{+\epsilon}dz ~~ E\vert\psi\rangle,
\end{equation}
it gives rise to a system of four linear coupled equations which are given by
\begin{subequations}
\begin{equation}
2v_z(b_0-b_\Delta) +i\tilde\Delta_z(b_0+b_\Delta)+i \left(\tilde{\Delta}_x-i \tilde{\Delta}_y\right)(d_0+d_\Delta)=0,
\end{equation}
\begin{equation}
2v_z(a_0-a_\Delta) + i\tilde\Delta_z(a_0+a_\Delta)+i \left(\tilde{\Delta}_x-i \tilde{\Delta}_y\right)(c_0+c_\Delta)=0,
\end{equation}
\begin{equation}
-2v_z(d_0-d_\Delta) + i\tilde\Delta_z(d_0+d_\Delta)-i \left(\tilde{\Delta}_x+i \tilde{\Delta}_y\right)(b_0+b_\Delta)=0,
\end{equation}
\begin{equation}
-2v_z(c_0-c_\Delta) + i\tilde\Delta_z(c_0+c_\Delta)-i \left(\tilde{\Delta}_x+i \tilde{\Delta}_y\right)(a_0+a_\Delta)=0.
\end{equation}
\end{subequations}
These equations can be rearranged to express the entries of $\vert \psi_\Delta\rangle$ in terms of the entries of $\vert\psi_0\rangle$ as
\begin{equation}
\begin{pmatrix}
a_\Delta\\b_\Delta\\c_\Delta\\ d_\Delta
\end{pmatrix} = \mathcal{M}\begin{pmatrix}
a_0\\b_0\\c_0\\ d_0
\end{pmatrix}
\end{equation}
where the matrix $\mathcal{M}$ is given by
\begin{equation}
\mathcal{M} = \frac{1}{m^2+\vert\bm\Delta\vert^2}
\begin{pmatrix}
 -\vert\bm\Delta\vert^2+m^2-2 i \Delta_z m & 0 & -2 \Delta_y m-2 \Delta_x i m & 0 \\
 0 & -\vert\bm\Delta\vert^2+m^2-2 i \Delta_z m & 0 & -2 \Delta_y m-2 \Delta_x i m \\
 2 \Delta_y m-2 i \Delta_x m & 0 & -\vert\bm\Delta\vert^2+m^2+2 \Delta_z i m & 0 \\
 0 & 2 \Delta_y m-2 i \Delta_x m & 0 & -\vert\bm\Delta\vert^2+m^2+2 \Delta_z i m \\
\end{pmatrix},
\end{equation}
which can be written in a simpler form given in Eq.\ref{eq:M}.

\twocolumngrid

\bibliography{references}

\begin{thebibliography}{42}%
\makeatletter
\providecommand \@ifxundefined [1]{%
 \@ifx{#1\undefined}
}%
\providecommand \@ifnum [1]{%
 \ifnum #1\expandafter \@firstoftwo
 \else \expandafter \@secondoftwo
 \fi
}%
\providecommand \@ifx [1]{%
 \ifx #1\expandafter \@firstoftwo
 \else \expandafter \@secondoftwo
 \fi
}%
\providecommand \natexlab [1]{#1}%
\providecommand \enquote  [1]{``#1''}%
\providecommand \bibnamefont  [1]{#1}%
\providecommand \bibfnamefont [1]{#1}%
\providecommand \citenamefont [1]{#1}%
\providecommand \href@noop [0]{\@secondoftwo}%
\providecommand \href [0]{\begingroup \@sanitize@url \@href}%
\providecommand \@href[1]{\@@startlink{#1}\@@href}%
\providecommand \@@href[1]{\endgroup#1\@@endlink}%
\providecommand \@sanitize@url [0]{\catcode `\\12\catcode `\$12\catcode
  `\&12\catcode `\#12\catcode `\^12\catcode `\_12\catcode `\%12\relax}%
\providecommand \@@startlink[1]{}%
\providecommand \@@endlink[0]{}%
\providecommand \url  [0]{\begingroup\@sanitize@url \@url }%
\providecommand \@url [1]{\endgroup\@href {#1}{\urlprefix }}%
\providecommand \urlprefix  [0]{URL }%
\providecommand \Eprint [0]{\href }%
\providecommand \doibase [0]{http://dx.doi.org/}%
\providecommand \selectlanguage [0]{\@gobble}%
\providecommand \bibinfo  [0]{\@secondoftwo}%
\providecommand \bibfield  [0]{\@secondoftwo}%
\providecommand \translation [1]{[#1]}%
\providecommand \BibitemOpen [0]{}%
\providecommand \bibitemStop [0]{}%
\providecommand \bibitemNoStop [0]{.\EOS\space}%
\providecommand \EOS [0]{\spacefactor3000\relax}%
\providecommand \BibitemShut  [1]{\csname bibitem#1\endcsname}%
\let\auto@bib@innerbib\@empty
%</preamble>
\bibitem [{\citenamefont {Kane}\ and\ \citenamefont
  {Mele}(2005{\natexlab{a}})}]{Kane2005}%
  \BibitemOpen
  \bibfield  {author} {\bibinfo {author} {\bibfnamefont {C.~L.}\ \bibnamefont
  {Kane}}\ and\ \bibinfo {author} {\bibfnamefont {E.~J.}\ \bibnamefont
  {Mele}},\ }\href {\doibase 10.1103/PhysRevLett.95.146802} {\bibfield
  {journal} {\bibinfo  {journal} {Phys. Rev. Lett.}\ }\textbf {\bibinfo
  {volume} {95}},\ \bibinfo {pages} {146802} (\bibinfo {year}
  {2005}{\natexlab{a}})}\BibitemShut {NoStop}%
\bibitem [{\citenamefont {Kane}\ and\ \citenamefont
  {Mele}(2005{\natexlab{b}})}]{Kane2005a}%
  \BibitemOpen
  \bibfield  {author} {\bibinfo {author} {\bibfnamefont {C.~L.}\ \bibnamefont
  {Kane}}\ and\ \bibinfo {author} {\bibfnamefont {E.~J.}\ \bibnamefont
  {Mele}},\ }\href {\doibase 10.1103/PhysRevLett.95.226801} {\bibfield
  {journal} {\bibinfo  {journal} {Phys. Rev. Lett.}\ }\textbf {\bibinfo
  {volume} {95}},\ \bibinfo {pages} {226801} (\bibinfo {year}
  {2005}{\natexlab{b}})}\BibitemShut {NoStop}%
\bibitem [{\citenamefont {Bernevig}\ and\ \citenamefont
  {Zhang}(2006)}]{Bernevig2006}%
  \BibitemOpen
  \bibfield  {author} {\bibinfo {author} {\bibfnamefont {B.~A.}\ \bibnamefont
  {Bernevig}}\ and\ \bibinfo {author} {\bibfnamefont {S.-C.}\ \bibnamefont
  {Zhang}},\ }\href {\doibase 10.1103/PhysRevLett.96.106802} {\bibfield
  {journal} {\bibinfo  {journal} {Phys. Rev. Lett.}\ }\textbf {\bibinfo
  {volume} {96}},\ \bibinfo {pages} {106802} (\bibinfo {year}
  {2006})}\BibitemShut {NoStop}%
\bibitem [{\citenamefont {Bernevig}\ \emph {et~al.}(2006)\citenamefont
  {Bernevig}, \citenamefont {Hughes},\ and\ \citenamefont
  {Zhang}}]{Bernevig2006a}%
  \BibitemOpen
  \bibfield  {author} {\bibinfo {author} {\bibfnamefont {B.~A.}\ \bibnamefont
  {Bernevig}}, \bibinfo {author} {\bibfnamefont {T.~L.}\ \bibnamefont
  {Hughes}}, \ and\ \bibinfo {author} {\bibfnamefont {S.-C.}\ \bibnamefont
  {Zhang}},\ }\href {\doibase 10.1126/science.1133734} {\bibfield  {journal}
  {\bibinfo  {journal} {Science (New York, N.Y.)}\ }\textbf {\bibinfo {volume}
  {314}},\ \bibinfo {pages} {1757} (\bibinfo {year} {2006})}\BibitemShut
  {NoStop}%
\bibitem [{\citenamefont {K\"{o}nig}\ \emph {et~al.}(2007)\citenamefont
  {K\"{o}nig}, \citenamefont {Wiedmann}, \citenamefont {Br\"{u}ne},
  \citenamefont {Roth}, \citenamefont {Buhmann}, \citenamefont {Molenkamp},
  \citenamefont {Qi},\ and\ \citenamefont {Zhang}}]{Konig2007}%
  \BibitemOpen
  \bibfield  {author} {\bibinfo {author} {\bibfnamefont {M.}~\bibnamefont
  {K\"{o}nig}}, \bibinfo {author} {\bibfnamefont {S.}~\bibnamefont {Wiedmann}},
  \bibinfo {author} {\bibfnamefont {C.}~\bibnamefont {Br\"{u}ne}}, \bibinfo
  {author} {\bibfnamefont {A.}~\bibnamefont {Roth}}, \bibinfo {author}
  {\bibfnamefont {H.}~\bibnamefont {Buhmann}}, \bibinfo {author} {\bibfnamefont
  {L.~W.}\ \bibnamefont {Molenkamp}}, \bibinfo {author} {\bibfnamefont {X.-L.}\
  \bibnamefont {Qi}}, \ and\ \bibinfo {author} {\bibfnamefont {S.-C.}\
  \bibnamefont {Zhang}},\ }\href {\doibase 10.1126/science.1148047} {\bibfield
  {journal} {\bibinfo  {journal} {Science (New York, N.Y.)}\ }\textbf {\bibinfo
  {volume} {318}},\ \bibinfo {pages} {766} (\bibinfo {year}
  {2007})}\BibitemShut {NoStop}%
\bibitem [{\citenamefont {Fu}\ \emph {et~al.}(2007)\citenamefont {Fu},
  \citenamefont {Kane},\ and\ \citenamefont {Mele}}]{Fu2007}%
  \BibitemOpen
  \bibfield  {author} {\bibinfo {author} {\bibfnamefont {L.}~\bibnamefont
  {Fu}}, \bibinfo {author} {\bibfnamefont {C.}~\bibnamefont {Kane}}, \ and\
  \bibinfo {author} {\bibfnamefont {E.}~\bibnamefont {Mele}},\ }\href {\doibase
  10.1103/PhysRevLett.98.106803} {\bibfield  {journal} {\bibinfo  {journal}
  {Phys. Rev. Lett.}\ }\textbf {\bibinfo {volume} {98}},\ \bibinfo {pages}
  {106803} (\bibinfo {year} {2007})}\BibitemShut {NoStop}%
\bibitem [{\citenamefont {Moore}(2010)}]{Moore2010}%
  \BibitemOpen
  \bibfield  {author} {\bibinfo {author} {\bibfnamefont {J.~E.}\ \bibnamefont
  {Moore}},\ }\href {\doibase 10.1038/nature08916} {\bibfield  {journal}
  {\bibinfo  {journal} {Nature}\ }\textbf {\bibinfo {volume} {464}},\ \bibinfo
  {pages} {194} (\bibinfo {year} {2010})}\BibitemShut {NoStop}%
\bibitem [{\citenamefont {Hasan}\ and\ \citenamefont {Kane}(2010)}]{Hasan2010}%
  \BibitemOpen
  \bibfield  {author} {\bibinfo {author} {\bibfnamefont {M.~Z.}\ \bibnamefont
  {Hasan}}\ and\ \bibinfo {author} {\bibfnamefont {C.~L.}\ \bibnamefont
  {Kane}},\ }\href {\doibase 10.1103/RevModPhys.82.3045} {\bibfield  {journal}
  {\bibinfo  {journal} {Reviews of Modern Physics}\ }\textbf {\bibinfo {volume}
  {82}},\ \bibinfo {pages} {3045} (\bibinfo {year} {2010})}\BibitemShut
  {NoStop}%
\bibitem [{\citenamefont {Qi}\ and\ \citenamefont {Zhang}(2011)}]{Qi2011}%
  \BibitemOpen
  \bibfield  {author} {\bibinfo {author} {\bibfnamefont {X.-L.}\ \bibnamefont
  {Qi}}\ and\ \bibinfo {author} {\bibfnamefont {S.-C.}\ \bibnamefont {Zhang}},\
  }\href {\doibase 10.1103/RevModPhys.83.1057} {\bibfield  {journal} {\bibinfo
  {journal} {Reviews of Modern Physics}\ }\textbf {\bibinfo {volume} {83}},\
  \bibinfo {pages} {1057} (\bibinfo {year} {2011})}\BibitemShut {NoStop}%
\bibitem [{\citenamefont {Bernevig}\ and\ \citenamefont
  {Hughes}(2013)}]{Bernevig2013book}%
  \BibitemOpen
  \bibfield  {author} {\bibinfo {author} {\bibfnamefont {B.}~\bibnamefont
  {Bernevig}}\ and\ \bibinfo {author} {\bibfnamefont {T.}~\bibnamefont
  {Hughes}},\ }\href {http://books.google.de/books?id=wOn7JHSSxrsC} {\emph
  {\bibinfo {title} {Topological Insulators and Topological Superconductors}}}\
  (\bibinfo  {publisher} {Princeton University Press},\ \bibinfo {year}
  {2013})\BibitemShut {NoStop}%
\bibitem [{\citenamefont {Zhang}\ \emph
  {et~al.}(2009{\natexlab{a}})\citenamefont {Zhang}, \citenamefont {Liu},
  \citenamefont {Qi}, \citenamefont {Dai}, \citenamefont {Fang},\ and\
  \citenamefont {Zhang}}]{Zhang2009}%
  \BibitemOpen
  \bibfield  {author} {\bibinfo {author} {\bibfnamefont {H.}~\bibnamefont
  {Zhang}}, \bibinfo {author} {\bibfnamefont {C.-X.}\ \bibnamefont {Liu}},
  \bibinfo {author} {\bibfnamefont {X.-L.}\ \bibnamefont {Qi}}, \bibinfo
  {author} {\bibfnamefont {X.}~\bibnamefont {Dai}}, \bibinfo {author}
  {\bibfnamefont {Z.}~\bibnamefont {Fang}}, \ and\ \bibinfo {author}
  {\bibfnamefont {S.-C.}\ \bibnamefont {Zhang}},\ }\href {\doibase
  10.1038/nphys1270} {\bibfield  {journal} {\bibinfo  {journal} {Nature
  Physics}\ }\textbf {\bibinfo {volume} {5}},\ \bibinfo {pages} {438} (\bibinfo
  {year} {2009}{\natexlab{a}})}\BibitemShut {NoStop}%
\bibitem [{\citenamefont {Chen}\ \emph {et~al.}(2009)\citenamefont {Chen},
  \citenamefont {Analytis}, \citenamefont {Chu}, \citenamefont {Liu},
  \citenamefont {Mo}, \citenamefont {Qi}, \citenamefont {Zhang}, \citenamefont
  {Lu}, \citenamefont {Dai}, \citenamefont {Fang}, \citenamefont {Zhang},
  \citenamefont {Fisher}, \citenamefont {Hussain},\ and\ \citenamefont
  {Shen}}]{Chen2009}%
  \BibitemOpen
  \bibfield  {author} {\bibinfo {author} {\bibfnamefont {Y.~L.}\ \bibnamefont
  {Chen}}, \bibinfo {author} {\bibfnamefont {J.~G.}\ \bibnamefont {Analytis}},
  \bibinfo {author} {\bibfnamefont {J.-H.}\ \bibnamefont {Chu}}, \bibinfo
  {author} {\bibfnamefont {Z.~K.}\ \bibnamefont {Liu}}, \bibinfo {author}
  {\bibfnamefont {S.-K.}\ \bibnamefont {Mo}}, \bibinfo {author} {\bibfnamefont
  {X.~L.}\ \bibnamefont {Qi}}, \bibinfo {author} {\bibfnamefont {H.~J.}\
  \bibnamefont {Zhang}}, \bibinfo {author} {\bibfnamefont {D.~H.}\ \bibnamefont
  {Lu}}, \bibinfo {author} {\bibfnamefont {X.}~\bibnamefont {Dai}}, \bibinfo
  {author} {\bibfnamefont {Z.}~\bibnamefont {Fang}}, \bibinfo {author}
  {\bibfnamefont {S.~C.}\ \bibnamefont {Zhang}}, \bibinfo {author}
  {\bibfnamefont {I.~R.}\ \bibnamefont {Fisher}}, \bibinfo {author}
  {\bibfnamefont {Z.}~\bibnamefont {Hussain}}, \ and\ \bibinfo {author}
  {\bibfnamefont {Z.-X.}\ \bibnamefont {Shen}},\ }\href {\doibase
  10.1126/science.1173034} {\bibfield  {journal} {\bibinfo  {journal} {Science
  (New York, N.Y.)}\ }\textbf {\bibinfo {volume} {325}},\ \bibinfo {pages}
  {178} (\bibinfo {year} {2009})}\BibitemShut {NoStop}%
\bibitem [{\citenamefont {Moore}\ and\ \citenamefont
  {Balents}(2007)}]{Moore2007}%
  \BibitemOpen
  \bibfield  {author} {\bibinfo {author} {\bibfnamefont {J.}~\bibnamefont
  {Moore}}\ and\ \bibinfo {author} {\bibfnamefont {L.}~\bibnamefont
  {Balents}},\ }\href {\doibase 10.1103/PhysRevB.75.121306} {\bibfield
  {journal} {\bibinfo  {journal} {Phys. Rev. B}\ }\textbf {\bibinfo {volume}
  {75}},\ \bibinfo {pages} {121306} (\bibinfo {year} {2007})}\BibitemShut
  {NoStop}%
\bibitem [{\citenamefont {Roy}(2009)}]{Roy2009}%
  \BibitemOpen
  \bibfield  {author} {\bibinfo {author} {\bibfnamefont {R.}~\bibnamefont
  {Roy}},\ }\href {\doibase 10.1103/PhysRevB.79.195322} {\bibfield  {journal}
  {\bibinfo  {journal} {Phys. Rev. B}\ }\textbf {\bibinfo {volume} {79}},\
  \bibinfo {pages} {195322} (\bibinfo {year} {2009})}\BibitemShut {NoStop}%
\bibitem [{\citenamefont {Qi}\ \emph {et~al.}(2008)\citenamefont {Qi},
  \citenamefont {Hughes},\ and\ \citenamefont {Zhang}}]{Qi2010}%
  \BibitemOpen
  \bibfield  {author} {\bibinfo {author} {\bibfnamefont {X.-L.}\ \bibnamefont
  {Qi}}, \bibinfo {author} {\bibfnamefont {T.~L.}\ \bibnamefont {Hughes}}, \
  and\ \bibinfo {author} {\bibfnamefont {S.-C.}\ \bibnamefont {Zhang}},\ }\href
  {\doibase 10.1103/PhysRevB.78.195424} {\bibfield  {journal} {\bibinfo
  {journal} {Phys. Rev. B}\ }\textbf {\bibinfo {volume} {78}},\ \bibinfo
  {pages} {195424} (\bibinfo {year} {2008})}\BibitemShut {NoStop}%
\bibitem [{\citenamefont {Hsieh}\ \emph {et~al.}(2008)\citenamefont {Hsieh},
  \citenamefont {Qian}, \citenamefont {Wray}, \citenamefont {Xia},
  \citenamefont {Hor}, \citenamefont {Cava},\ and\ \citenamefont
  {Hasan}}]{Hsieh2008}%
  \BibitemOpen
  \bibfield  {author} {\bibinfo {author} {\bibfnamefont {D.}~\bibnamefont
  {Hsieh}}, \bibinfo {author} {\bibfnamefont {D.}~\bibnamefont {Qian}},
  \bibinfo {author} {\bibfnamefont {L.}~\bibnamefont {Wray}}, \bibinfo {author}
  {\bibfnamefont {Y.}~\bibnamefont {Xia}}, \bibinfo {author} {\bibfnamefont
  {Y.~S.}\ \bibnamefont {Hor}}, \bibinfo {author} {\bibfnamefont {R.~J.}\
  \bibnamefont {Cava}}, \ and\ \bibinfo {author} {\bibfnamefont {M.~Z.}\
  \bibnamefont {Hasan}},\ }\href {\doibase 10.1038/nature06843} {\bibfield
  {journal} {\bibinfo  {journal} {Nature}\ }\textbf {\bibinfo {volume} {452}},\
  \bibinfo {pages} {970} (\bibinfo {year} {2008})}\BibitemShut {NoStop}%
\bibitem [{\citenamefont {Xia}\ \emph {et~al.}(2009)\citenamefont {Xia},
  \citenamefont {Qian}, \citenamefont {Hsieh}, \citenamefont {Wray},
  \citenamefont {Pal}, \citenamefont {Lin}, \citenamefont {Bansil},
  \citenamefont {Grauer}, \citenamefont {Hor}, \citenamefont {Cava},\ and\
  \citenamefont {Hasan}}]{Xia2009}%
  \BibitemOpen
  \bibfield  {author} {\bibinfo {author} {\bibfnamefont {Y.}~\bibnamefont
  {Xia}}, \bibinfo {author} {\bibfnamefont {D.}~\bibnamefont {Qian}}, \bibinfo
  {author} {\bibfnamefont {D.}~\bibnamefont {Hsieh}}, \bibinfo {author}
  {\bibfnamefont {L.}~\bibnamefont {Wray}}, \bibinfo {author} {\bibfnamefont
  {a.}~\bibnamefont {Pal}}, \bibinfo {author} {\bibfnamefont {H.}~\bibnamefont
  {Lin}}, \bibinfo {author} {\bibfnamefont {A.}~\bibnamefont {Bansil}},
  \bibinfo {author} {\bibfnamefont {D.}~\bibnamefont {Grauer}}, \bibinfo
  {author} {\bibfnamefont {Y.~S.}\ \bibnamefont {Hor}}, \bibinfo {author}
  {\bibfnamefont {R.~J.}\ \bibnamefont {Cava}}, \ and\ \bibinfo {author}
  {\bibfnamefont {M.~Z.}\ \bibnamefont {Hasan}},\ }\href {\doibase
  10.1038/nphys1274} {\bibfield  {journal} {\bibinfo  {journal} {Nature
  Physics}\ }\textbf {\bibinfo {volume} {5}},\ \bibinfo {pages} {398} (\bibinfo
  {year} {2009})}\BibitemShut {NoStop}%
\bibitem [{\citenamefont {Hasan}\ \emph {et~al.}(2014)\citenamefont {Hasan},
  \citenamefont {Xu}, \citenamefont {Hsieh}, \citenamefont {Wray},\ and\
  \citenamefont {Xia}}]{Hasan2014}%
  \BibitemOpen
  \bibfield  {author} {\bibinfo {author} {\bibfnamefont {M.~Z.}\ \bibnamefont
  {Hasan}}, \bibinfo {author} {\bibfnamefont {S.-Y.}\ \bibnamefont {Xu}},
  \bibinfo {author} {\bibfnamefont {D.}~\bibnamefont {Hsieh}}, \bibinfo
  {author} {\bibfnamefont {L.~A.}\ \bibnamefont {Wray}}, \ and\ \bibinfo
  {author} {\bibfnamefont {Y.}~\bibnamefont {Xia}},\ }\href
  {http://arxiv.org/abs/1401.0848v1} {\  (\bibinfo {year} {2014})},\ \Eprint
  {http://arxiv.org/abs/1401.0848} {arXiv:1401.0848} \BibitemShut {NoStop}%
\bibitem [{\citenamefont {Roushan}\ \emph {et~al.}(2009)\citenamefont
  {Roushan}, \citenamefont {Seo}, \citenamefont {Parker}, \citenamefont {Hor},
  \citenamefont {Hsieh}, \citenamefont {Qian}, \citenamefont {Richardella},
  \citenamefont {Hasan}, \citenamefont {Cava},\ and\ \citenamefont
  {Yazdani}}]{Roushan2009}%
  \BibitemOpen
  \bibfield  {author} {\bibinfo {author} {\bibfnamefont {P.}~\bibnamefont
  {Roushan}}, \bibinfo {author} {\bibfnamefont {J.}~\bibnamefont {Seo}},
  \bibinfo {author} {\bibfnamefont {C.~V.}\ \bibnamefont {Parker}}, \bibinfo
  {author} {\bibfnamefont {Y.~S.}\ \bibnamefont {Hor}}, \bibinfo {author}
  {\bibfnamefont {D.}~\bibnamefont {Hsieh}}, \bibinfo {author} {\bibfnamefont
  {D.}~\bibnamefont {Qian}}, \bibinfo {author} {\bibfnamefont {A.}~\bibnamefont
  {Richardella}}, \bibinfo {author} {\bibfnamefont {M.~Z.}\ \bibnamefont
  {Hasan}}, \bibinfo {author} {\bibfnamefont {R.~J.}\ \bibnamefont {Cava}}, \
  and\ \bibinfo {author} {\bibfnamefont {A.}~\bibnamefont {Yazdani}},\ }\href
  {\doibase 10.1038/nature08308} {\bibfield  {journal} {\bibinfo  {journal}
  {Nature}\ }\textbf {\bibinfo {volume} {460}},\ \bibinfo {pages} {1106}
  (\bibinfo {year} {2009})}\BibitemShut {NoStop}%
\bibitem [{\citenamefont {Zhang}\ \emph
  {et~al.}(2009{\natexlab{b}})\citenamefont {Zhang}, \citenamefont {Cheng},
  \citenamefont {Chen}, \citenamefont {Jia}, \citenamefont {Ma}, \citenamefont
  {He}, \citenamefont {Wang}, \citenamefont {Zhang}, \citenamefont {Dai},
  \citenamefont {Fang}, \citenamefont {Xie},\ and\ \citenamefont
  {Xue}}]{Zhang2009a}%
  \BibitemOpen
  \bibfield  {author} {\bibinfo {author} {\bibfnamefont {T.}~\bibnamefont
  {Zhang}}, \bibinfo {author} {\bibfnamefont {P.}~\bibnamefont {Cheng}},
  \bibinfo {author} {\bibfnamefont {X.}~\bibnamefont {Chen}}, \bibinfo {author}
  {\bibfnamefont {J.-F.}\ \bibnamefont {Jia}}, \bibinfo {author} {\bibfnamefont
  {X.}~\bibnamefont {Ma}}, \bibinfo {author} {\bibfnamefont {K.}~\bibnamefont
  {He}}, \bibinfo {author} {\bibfnamefont {L.}~\bibnamefont {Wang}}, \bibinfo
  {author} {\bibfnamefont {H.}~\bibnamefont {Zhang}}, \bibinfo {author}
  {\bibfnamefont {X.}~\bibnamefont {Dai}}, \bibinfo {author} {\bibfnamefont
  {Z.}~\bibnamefont {Fang}}, \bibinfo {author} {\bibfnamefont {X.}~\bibnamefont
  {Xie}}, \ and\ \bibinfo {author} {\bibfnamefont {Q.-K.}\ \bibnamefont
  {Xue}},\ }\href {\doibase 10.1103/PhysRevLett.103.266803} {\bibfield
  {journal} {\bibinfo  {journal} {Phys. Rev. Lett.}\ }\textbf {\bibinfo
  {volume} {103}},\ \bibinfo {pages} {266803} (\bibinfo {year}
  {2009}{\natexlab{b}})}\BibitemShut {NoStop}%
\bibitem [{\citenamefont {Alpichshev}\ \emph {et~al.}(2010)\citenamefont
  {Alpichshev}, \citenamefont {Analytis}, \citenamefont {Chu}, \citenamefont
  {Fisher}, \citenamefont {Chen}, \citenamefont {Shen}, \citenamefont {Fang},\
  and\ \citenamefont {Kapitulnik}}]{Alpichshev2010}%
  \BibitemOpen
  \bibfield  {author} {\bibinfo {author} {\bibfnamefont {Z.}~\bibnamefont
  {Alpichshev}}, \bibinfo {author} {\bibfnamefont {J.~G.}\ \bibnamefont
  {Analytis}}, \bibinfo {author} {\bibfnamefont {J.-H.}\ \bibnamefont {Chu}},
  \bibinfo {author} {\bibfnamefont {I.~R.}\ \bibnamefont {Fisher}}, \bibinfo
  {author} {\bibfnamefont {Y.~L.}\ \bibnamefont {Chen}}, \bibinfo {author}
  {\bibfnamefont {Z.~X.}\ \bibnamefont {Shen}}, \bibinfo {author}
  {\bibfnamefont {A.}~\bibnamefont {Fang}}, \ and\ \bibinfo {author}
  {\bibfnamefont {A.}~\bibnamefont {Kapitulnik}},\ }\href {\doibase
  10.1103/PhysRevLett.104.016401} {\bibfield  {journal} {\bibinfo  {journal}
  {Phys. Rev. Lett.}\ }\textbf {\bibinfo {volume} {104}},\ \bibinfo {pages}
  {016401} (\bibinfo {year} {2010})}\BibitemShut {NoStop}%
\bibitem [{\citenamefont {Yokoyama}\ \emph {et~al.}(2010)\citenamefont
  {Yokoyama}, \citenamefont {Tanaka},\ and\ \citenamefont
  {Nagaosa}}]{Yokoyama2010}%
  \BibitemOpen
  \bibfield  {author} {\bibinfo {author} {\bibfnamefont {T.}~\bibnamefont
  {Yokoyama}}, \bibinfo {author} {\bibfnamefont {Y.}~\bibnamefont {Tanaka}}, \
  and\ \bibinfo {author} {\bibfnamefont {N.}~\bibnamefont {Nagaosa}},\ }\href
  {\doibase 10.1103/PhysRevB.81.121401} {\bibfield  {journal} {\bibinfo
  {journal} {Phys. Rev. B}\ }\textbf {\bibinfo {volume} {81}},\ \bibinfo
  {pages} {121401} (\bibinfo {year} {2010})}\BibitemShut {NoStop}%
\bibitem [{\citenamefont {Taguchi}\ \emph {et~al.}(2014)\citenamefont
  {Taguchi}, \citenamefont {Yokoyama},\ and\ \citenamefont
  {Tanaka}}]{Taguchi2014}%
  \BibitemOpen
  \bibfield  {author} {\bibinfo {author} {\bibfnamefont {K.}~\bibnamefont
  {Taguchi}}, \bibinfo {author} {\bibfnamefont {T.}~\bibnamefont {Yokoyama}}, \
  and\ \bibinfo {author} {\bibfnamefont {Y.}~\bibnamefont {Tanaka}},\ }\href
  {\doibase 10.1103/PhysRevB.89.085407} {\bibfield  {journal} {\bibinfo
  {journal} {Phys. Rev. B}\ }\textbf {\bibinfo {volume} {89}},\ \bibinfo
  {pages} {085407} (\bibinfo {year} {2014})}\BibitemShut {NoStop}%
\bibitem [{\citenamefont {Li}\ \emph {et~al.}(2014)\citenamefont {Li},
  \citenamefont {{van 't Erve}}, \citenamefont {Robinson}, \citenamefont {Liu},
  \citenamefont {Li},\ and\ \citenamefont {Jonker}}]{Li2014a}%
  \BibitemOpen
  \bibfield  {author} {\bibinfo {author} {\bibfnamefont {C.~H.}\ \bibnamefont
  {Li}}, \bibinfo {author} {\bibfnamefont {O.~M.~J.}\ \bibnamefont {{van 't
  Erve}}}, \bibinfo {author} {\bibfnamefont {J.~T.}\ \bibnamefont {Robinson}},
  \bibinfo {author} {\bibfnamefont {Y.}~\bibnamefont {Liu}}, \bibinfo {author}
  {\bibfnamefont {L.}~\bibnamefont {Li}}, \ and\ \bibinfo {author}
  {\bibfnamefont {B.~T.}\ \bibnamefont {Jonker}},\ }\href {\doibase
  10.1038/nnano.2014.16} {\bibfield  {journal} {\bibinfo  {journal} {Nature
  nanotechnology}\ }\textbf {\bibinfo {volume} {9}},\ \bibinfo {pages} {218}
  (\bibinfo {year} {2014})}\BibitemShut {NoStop}%
\bibitem [{\citenamefont {Liu}\ \emph {et~al.}()\citenamefont {Liu},
  \citenamefont {Richardella}, \citenamefont {Garate},\ and\ \citenamefont
  {Zhu}}]{Liu2014}%
  \BibitemOpen
  \bibfield  {author} {\bibinfo {author} {\bibfnamefont {L.}~\bibnamefont
  {Liu}}, \bibinfo {author} {\bibfnamefont {A.}~\bibnamefont {Richardella}},
  \bibinfo {author} {\bibfnamefont {I.}~\bibnamefont {Garate}}, \ and\ \bibinfo
  {author} {\bibfnamefont {Y.}~\bibnamefont {Zhu}},\ }\href
  {http://scholar.google.com/scholar?hl=en\&btnG=Search\&q=intitle:No+Title\#0
  http://arxiv.org/abs/1410.7494} {\ }\Eprint {http://arxiv.org/abs/1410.7494}
  {arXiv:1410.7494} \BibitemShut {NoStop}%
\bibitem [{\citenamefont {Dankert}\ \emph {et~al.}(2014)\citenamefont
  {Dankert}, \citenamefont {Geurs}, \citenamefont {Kamalakar},\ and\
  \citenamefont {Dash}}]{Dankert2014}%
  \BibitemOpen
  \bibfield  {author} {\bibinfo {author} {\bibfnamefont {A.}~\bibnamefont
  {Dankert}}, \bibinfo {author} {\bibfnamefont {J.}~\bibnamefont {Geurs}},
  \bibinfo {author} {\bibfnamefont {M.}~\bibnamefont {Kamalakar}}, \ and\
  \bibinfo {author} {\bibfnamefont {S.}~\bibnamefont {Dash}},\ }\href
  {http://arxiv.org/abs/1410.8038} {\  (\bibinfo {year} {2014})},\ \Eprint
  {http://arxiv.org/abs/1410.8038v2} {arXiv:1410.8038v2} \BibitemShut {NoStop}%
\bibitem [{\citenamefont {Roy}\ \emph {et~al.}(2015{\natexlab{a}})\citenamefont
  {Roy}, \citenamefont {Soori},\ and\ \citenamefont {Das}}]{Roy2015}%
  \BibitemOpen
  \bibfield  {author} {\bibinfo {author} {\bibfnamefont {S.}~\bibnamefont
  {Roy}}, \bibinfo {author} {\bibfnamefont {A.}~\bibnamefont {Soori}}, \ and\
  \bibinfo {author} {\bibfnamefont {S.}~\bibnamefont {Das}},\ }\href {\doibase
  10.1103/PhysRevB.91.041109} {\bibfield  {journal} {\bibinfo  {journal} {Phys.
  Rev. B}\ }\textbf {\bibinfo {volume} {91}},\ \bibinfo {pages} {041109}
  (\bibinfo {year} {2015}{\natexlab{a}})}\BibitemShut {NoStop}%
\bibitem [{\citenamefont {Roy}\ \emph {et~al.}(2015{\natexlab{b}})\citenamefont
  {Roy}, \citenamefont {Saha},\ and\ \citenamefont {Das}}]{Roy2015a}%
  \BibitemOpen
  \bibfield  {author} {\bibinfo {author} {\bibfnamefont {S.}~\bibnamefont
  {Roy}}, \bibinfo {author} {\bibfnamefont {K.}~\bibnamefont {Saha}}, \ and\
  \bibinfo {author} {\bibfnamefont {S.}~\bibnamefont {Das}},\ }\href {\doibase
  10.1103/PhysRevB.91.195415} {\bibfield  {journal} {\bibinfo  {journal} {Phys.
  Rev. B}\ }\textbf {\bibinfo {volume} {91}},\ \bibinfo {pages} {195415}
  (\bibinfo {year} {2015}{\natexlab{b}})}\BibitemShut {NoStop}%
\bibitem [{\citenamefont {Slonczewski}(1989)}]{Slonczewski1989}%
  \BibitemOpen
  \bibfield  {author} {\bibinfo {author} {\bibfnamefont {J.}~\bibnamefont
  {Slonczewski}},\ }\href
  {http://journals.aps.org/prb/abstract/10.1103/PhysRevB.39.6995} {\bibfield
  {journal} {\bibinfo  {journal} {Phys. Rev. B}\ }\textbf {\bibinfo {volume}
  {39}},\ \bibinfo {pages} {6995} (\bibinfo {year} {1989})}\BibitemShut
  {NoStop}%
\bibitem [{\citenamefont {\ifmmode \check{Z}\else
  \v{Z}\fi{}uti\ifmmode~\acute{c}\else \'{c}\fi{}}\ \emph
  {et~al.}(2004)\citenamefont {\ifmmode \check{Z}\else
  \v{Z}\fi{}uti\ifmmode~\acute{c}\else \'{c}\fi{}}, \citenamefont {Fabian},\
  and\ \citenamefont {Das~Sarma}}]{Zutic2004}%
  \BibitemOpen
  \bibfield  {author} {\bibinfo {author} {\bibfnamefont {I.}~\bibnamefont
  {\ifmmode \check{Z}\else \v{Z}\fi{}uti\ifmmode~\acute{c}\else \'{c}\fi{}}},
  \bibinfo {author} {\bibfnamefont {J.}~\bibnamefont {Fabian}}, \ and\ \bibinfo
  {author} {\bibfnamefont {S.}~\bibnamefont {Das~Sarma}},\ }\href {\doibase
  10.1103/RevModPhys.76.323} {\bibfield  {journal} {\bibinfo  {journal} {Rev.
  Mod. Phys.}\ }\textbf {\bibinfo {volume} {76}},\ \bibinfo {pages} {323}
  (\bibinfo {year} {2004})}\BibitemShut {NoStop}%
\bibitem [{\citenamefont {Yokoyama}\ and\ \citenamefont
  {Murakami}(2014)}]{Yokoyama2014}%
  \BibitemOpen
  \bibfield  {author} {\bibinfo {author} {\bibfnamefont {T.}~\bibnamefont
  {Yokoyama}}\ and\ \bibinfo {author} {\bibfnamefont {S.}~\bibnamefont
  {Murakami}},\ }\href {\doibase 10.1016/j.physe.2013.07.015} {\bibfield
  {journal} {\bibinfo  {journal} {Physica E: Low-dimensional Systems and
  Nanostructures}\ }\textbf {\bibinfo {volume} {55}},\ \bibinfo {pages} {1}
  (\bibinfo {year} {2014})}\BibitemShut {NoStop}%
\bibitem [{\citenamefont {Zhang}\ \emph {et~al.}(2012)\citenamefont {Zhang},
  \citenamefont {Kane},\ and\ \citenamefont {Mele}}]{Zhang2012}%
  \BibitemOpen
  \bibfield  {author} {\bibinfo {author} {\bibfnamefont {F.}~\bibnamefont
  {Zhang}}, \bibinfo {author} {\bibfnamefont {C.~L.}\ \bibnamefont {Kane}}, \
  and\ \bibinfo {author} {\bibfnamefont {E.~J.}\ \bibnamefont {Mele}},\ }\href
  {\doibase 10.1103/PhysRevB.86.081303} {\bibfield  {journal} {\bibinfo
  {journal} {Phys. Rev. B}\ }\textbf {\bibinfo {volume} {86}},\ \bibinfo
  {pages} {081303} (\bibinfo {year} {2012})}\BibitemShut {NoStop}%
\bibitem [{\citenamefont {Liu}\ \emph {et~al.}(2010)\citenamefont {Liu},
  \citenamefont {Qi}, \citenamefont {Zhang}, \citenamefont {Dai}, \citenamefont
  {Fang},\ and\ \citenamefont {Zhang}}]{Liu2010}%
  \BibitemOpen
  \bibfield  {author} {\bibinfo {author} {\bibfnamefont {C.-X.}\ \bibnamefont
  {Liu}}, \bibinfo {author} {\bibfnamefont {X.-L.}\ \bibnamefont {Qi}},
  \bibinfo {author} {\bibfnamefont {H.}~\bibnamefont {Zhang}}, \bibinfo
  {author} {\bibfnamefont {X.}~\bibnamefont {Dai}}, \bibinfo {author}
  {\bibfnamefont {Z.}~\bibnamefont {Fang}}, \ and\ \bibinfo {author}
  {\bibfnamefont {S.-C.}\ \bibnamefont {Zhang}},\ }\href {\doibase
  10.1103/PhysRevB.82.045122} {\bibfield  {journal} {\bibinfo  {journal} {Phys.
  Rev. B}\ }\textbf {\bibinfo {volume} {82}},\ \bibinfo {pages} {045122}
  (\bibinfo {year} {2010})}\BibitemShut {NoStop}%
\bibitem [{\citenamefont {Jackiw}\ and\ \citenamefont
  {Rebbi}(1976)}]{Jackiw1976}%
  \BibitemOpen
  \bibfield  {author} {\bibinfo {author} {\bibfnamefont {R.}~\bibnamefont
  {Jackiw}}\ and\ \bibinfo {author} {\bibfnamefont {C.}~\bibnamefont {Rebbi}},\
  }\href {\doibase 10.1103/PhysRevD.13.3398} {\bibfield  {journal} {\bibinfo
  {journal} {Phys. Rev. D}\ }\textbf {\bibinfo {volume} {13}},\ \bibinfo
  {pages} {3398} (\bibinfo {year} {1976})}\BibitemShut {NoStop}%
\bibitem [{Note1()}]{Note1}%
  \BibitemOpen
  \bibinfo {note} {The optimization used by the authors to verify the protocol
  was implemented using the Broyden-Fletcher-Goldfarb-Shanno algorithm which
  gave relative errors of magnitudes less than $10^{-8}$. More details of the
  algorithm can be found in Ref.\protect \rev@citealpnum
  {Nocedal2006}}\BibitemShut {NoStop}%
\bibitem [{\citenamefont {Wiesendanger}(2009)}]{Wiesendanger2009}%
  \BibitemOpen
  \bibfield  {author} {\bibinfo {author} {\bibfnamefont {R.}~\bibnamefont
  {Wiesendanger}},\ }\href {\doibase 10.1103/RevModPhys.81.1495} {\bibfield
  {journal} {\bibinfo  {journal} {Rev. Mod. Phys.}\ }\textbf {\bibinfo {volume}
  {81}},\ \bibinfo {pages} {1495} (\bibinfo {year} {2009})}\BibitemShut
  {NoStop}%
\bibitem [{\citenamefont {Wachowiak}\ \emph {et~al.}(2002)\citenamefont
  {Wachowiak}, \citenamefont {Wiebe}, \citenamefont {Bode}, \citenamefont
  {Pietzsch}, \citenamefont {Morgenstern},\ and\ \citenamefont
  {Wiesendanger}}]{Wachowiak2002}%
  \BibitemOpen
  \bibfield  {author} {\bibinfo {author} {\bibfnamefont {A.}~\bibnamefont
  {Wachowiak}}, \bibinfo {author} {\bibfnamefont {J.}~\bibnamefont {Wiebe}},
  \bibinfo {author} {\bibfnamefont {M.}~\bibnamefont {Bode}}, \bibinfo {author}
  {\bibfnamefont {O.}~\bibnamefont {Pietzsch}}, \bibinfo {author}
  {\bibfnamefont {M.}~\bibnamefont {Morgenstern}}, \ and\ \bibinfo {author}
  {\bibfnamefont {R.}~\bibnamefont {Wiesendanger}},\ }\href@noop {} {\bibfield
  {journal} {\bibinfo  {journal} {Science}\ }\textbf {\bibinfo {volume}
  {298}},\ \bibinfo {pages} {577} (\bibinfo {year} {2002})}\BibitemShut
  {NoStop}%
\bibitem [{\citenamefont {Yang}\ \emph {et~al.}(2013)\citenamefont {Yang},
  \citenamefont {Song}, \citenamefont {Li}, \citenamefont {Zhang},
  \citenamefont {Yao}, \citenamefont {Liu}, \citenamefont {Qian}, \citenamefont
  {Gao},\ and\ \citenamefont {Jia}}]{Yang2013}%
  \BibitemOpen
  \bibfield  {author} {\bibinfo {author} {\bibfnamefont {F.}~\bibnamefont
  {Yang}}, \bibinfo {author} {\bibfnamefont {Y.~R.}\ \bibnamefont {Song}},
  \bibinfo {author} {\bibfnamefont {H.}~\bibnamefont {Li}}, \bibinfo {author}
  {\bibfnamefont {K.~F.}\ \bibnamefont {Zhang}}, \bibinfo {author}
  {\bibfnamefont {X.}~\bibnamefont {Yao}}, \bibinfo {author} {\bibfnamefont
  {C.}~\bibnamefont {Liu}}, \bibinfo {author} {\bibfnamefont {D.}~\bibnamefont
  {Qian}}, \bibinfo {author} {\bibfnamefont {C.~L.}\ \bibnamefont {Gao}}, \
  and\ \bibinfo {author} {\bibfnamefont {J.-F.}\ \bibnamefont {Jia}},\ }\href
  {\doibase 10.1103/PhysRevLett.111.176802} {\bibfield  {journal} {\bibinfo
  {journal} {Phys. Rev. Lett.}\ }\textbf {\bibinfo {volume} {111}},\ \bibinfo
  {pages} {176802} (\bibinfo {year} {2013})}\BibitemShut {NoStop}%
\bibitem [{\citenamefont {Pesin}\ and\ \citenamefont
  {MacDonald}(2012)}]{Pesin2012}%
  \BibitemOpen
  \bibfield  {author} {\bibinfo {author} {\bibfnamefont {D.}~\bibnamefont
  {Pesin}}\ and\ \bibinfo {author} {\bibfnamefont {A.~H.}\ \bibnamefont
  {MacDonald}},\ }\href@noop {} {\bibfield  {journal} {\bibinfo  {journal}
  {Nature materials}\ }\textbf {\bibinfo {volume} {11}},\ \bibinfo {pages}
  {409} (\bibinfo {year} {2012})}\BibitemShut {NoStop}%
\bibitem [{\citenamefont {Qu}\ \emph {et~al.}(2010)\citenamefont {Qu},
  \citenamefont {Hor}, \citenamefont {Xiong}, \citenamefont {Cava},\ and\
  \citenamefont {Ong}}]{Qu2010}%
  \BibitemOpen
  \bibfield  {author} {\bibinfo {author} {\bibfnamefont {D.-X.}\ \bibnamefont
  {Qu}}, \bibinfo {author} {\bibfnamefont {Y.}~\bibnamefont {Hor}}, \bibinfo
  {author} {\bibfnamefont {J.}~\bibnamefont {Xiong}}, \bibinfo {author}
  {\bibfnamefont {R.}~\bibnamefont {Cava}}, \ and\ \bibinfo {author}
  {\bibfnamefont {N.}~\bibnamefont {Ong}},\ }\href@noop {} {\bibfield
  {journal} {\bibinfo  {journal} {Science}\ }\textbf {\bibinfo {volume}
  {329}},\ \bibinfo {pages} {821} (\bibinfo {year} {2010})}\BibitemShut
  {NoStop}%
\bibitem [{\citenamefont {Fu}(2009)}]{Fu2009}%
  \BibitemOpen
  \bibfield  {author} {\bibinfo {author} {\bibfnamefont {L.}~\bibnamefont
  {Fu}},\ }\href {\doibase 10.1103/PhysRevLett.103.266801} {\bibfield
  {journal} {\bibinfo  {journal} {Phys. Rev. Lett.}\ }\textbf {\bibinfo
  {volume} {103}},\ \bibinfo {pages} {266801} (\bibinfo {year}
  {2009})}\BibitemShut {NoStop}%
\bibitem [{\citenamefont {Nocedal}\ and\ \citenamefont
  {Wright}(2006)}]{Nocedal2006}%
  \BibitemOpen
  \bibfield  {author} {\bibinfo {author} {\bibfnamefont {J.}~\bibnamefont
  {Nocedal}}\ and\ \bibinfo {author} {\bibfnamefont {S.}~\bibnamefont
  {Wright}},\ }\href@noop {} {\emph {\bibinfo {title} {Numerical
  optimization}}}\ (\bibinfo  {publisher} {Springer Science \& Business
  Media},\ \bibinfo {year} {2006})\BibitemShut {NoStop}%
\end{thebibliography}%

\end{document}